\documentclass[graybox, secnum]{svmult}

\usepackage[american]{babel}
\usepackage{amsfonts}
\usepackage{amsmath}
\usepackage{bbold}
\usepackage{amssymb}
\usepackage{epsfig}
\usepackage{latexsym}
\usepackage{paralist}
\usepackage{fancyhdr}
\usepackage{graphicx}
\usepackage[vcentermath]{youngtab}
\usepackage{young}
\usepackage{ytableau}
\usepackage{etex}
\usepackage{braket}
\usepackage{float}
\usepackage{slashed}
\usepackage{xcolor}
\usepackage{soul}
\usepackage{dcolumn} 
\usepackage{physics}
\usepackage{bm}
\usepackage{nicefrac}
\usepackage[toc,page]{appendix}

\usepackage{xfp}

\usepackage{mathptmx}       % selects Times Roman as basic font
\usepackage{helvet}         % selects Helvetica as sans-serif font
\usepackage{courier}        % selects Courier as typewriter font
\usepackage{type1cm}        % activate if the above 3 fonts are
\usepackage{makeidx}         % allows index generation
\usepackage{graphicx}        % standard LaTeX graphics tool
\usepackage{multicol}        % used for the two-column index
\usepackage[bottom]{footmisc}% places footnotes at page bottom
\usepackage{hyperref}        %for hyperlinks
\usepackage{soul}            % for high-lighting of text
\hypersetup{colorlinks=true,urlcolor=blue}

\usepackage[square,numbers]{natbib}

\makeindex             % used for the subject index
\begin{document}

%\tableofcontents{}


\title*{Perturbative approaches to non-perturbative quantum gravity}

\author{Riccardo Martini\thanks{corresponding author}, Gian Paolo Vacca and Omar Zanusso}

\institute{
Riccardo Martini
\at
INFN - Sezione di Pisa, Largo Bruno Pontecorvo 3, 56127 Pisa, Italy \email{riccardo.martini@pi.infn.it}
\and
Gian Paolo Vacca
\at
INFN - Sezione di Bologna, via Irnerio 46, 40126 Bologna, Italy, \email{vacca@bo.infn.it}
\and
Omar Zanusso
\at
Universit\`a di Pisa and INFN - Sezione di Pisa, Largo Bruno Pontecorvo 3, 56127 Pisa, Italy \email{omar.zanusso@unipi.it}
}


\maketitle
\abstract{
We discuss the birth of the nonperturbative approach to quantum gravity known as
quantum Einstein gravity, in which the gravitational interactions are conjectured to be asymptotically safe. The interactions are assumed to be finite and consistent at high energies thanks to a scale-invariant ultraviolet completion.
We present the framework on the basis of perturbative arguments that originally motivated it, paying special attention to the $\epsilon$-expansion in $d=2+\epsilon$ dimensions and the large-$N$ expansion for $N$ the number of flavors of matter fields.
The chapter is organized in such a way that each section is mostly independent and can offer several ideas for both conceptual and technical future developments.
}

\section*{Keywords} 
Quantum gravity, asymptotic safety, UV completion, perturbation theory, analytic continuation

%%%%%%%%%%%%%%%%%%%%%%%%%%%%%%%%%%%%%%%%%%%%%%%%%%%%%%%%%%%%%
%%%%%%%%%%%%%%%%%%%%%%%%%%%%%%%%%%%%%%%%%%%%%%%%%%%%%%%%%%%%%
\section{Introduction}
%%%%%%%%%%%%%%%%%%%%%%%%%%%%%%%%%%%%%%%%%%%%%%%%%%%%%%%%%%%%%
%%%%%%%%%%%%%%%%%%%%%%%%%%%%%%%%%%%%%%%%%%%%%%%%%%%%%%%%%%%%%

The {\it low energy} description of the gravitational interactions at human, astrophysical and cosmological scales has been successfully encoded in the classical field theory of general relativity, which adopts the metric field as the main ingredient with a redundant gauge formulation based on the group of diffeomorphisms.
There is, in principle, the need to reconcile this description with our understanding of the quantum world, which contains quantum matter,
and, eventually, obtain a quantum theory of the gravitational interactions.
The simplest idea in this direction would be to promote the metric field and the Einstein-Hilbert action to a path-integral and check
if this lift makes sense.
Here we are under the assumption that any meaningful canonical approach to gravity's quantization, for example in terms of a wave functional of a metric, could be constructed only provided an ultraviolet complete description is available, or else it would not work at arbitrarily high energies.

It has been well-known since a long time ago that (the path-integral of) Einstein-Hilbert gravity is not \emph{perturbatively} renormalizable, because it leads to on-shell nonrenormalizable divergences that appear at two-loops for pure gravity \cite{
Goroff:1985sz, Goroff:1985th,vandeVen:1991gw}, and at one-loop for gravity
with a cosmological constant \cite{Christensen:1979iy} or gravity
coupled to matter \cite{Barvinsky:1993zg,Deser:1974cy} and gauge fields \cite{Deser:1974zzd}. From a field-theoretical point of view, the implication is that the weakly-coupled Gaussian limit of Einstein's metric gravity cannot be the ultraviolet limit of a theory of quantum gravity, in stark contrast to what happens to asymptotically free theories such as Yang-Mills gauge theories for certain gauge groups and number of flavor fields. Historically, this understanding has spawned a great deal of work in alternative approaches
in which either the fundamental degrees of freedom were changed (e.g., string theory) or the methods were revisited (e.g., loop quantum gravity).
The early days of metric's quantum gravity are covered in Sect.~\ref{sec:falureRen} of this chapter, starting from the seminal work of 't Hooft \& Veltman \cite{tHooft:1974toh}.

A less-known fact is that the construction of a \emph{low-energy effective} quantum theory of gravity by standard path-integral methods is still meaningful if the loop expansion in the (reduced) Planck's constant $\hbar$ is replaced by an effective expansion in some mass scale, which is generally identified with Planck's mass $M_{\rm Pl}$ \cite{Donoghue:1994dn,Burgess:2003jk}. In this case, the ``effects'' of the previously unwanted divergences of perturbation theory are suppressed by powers of $E/M_{\rm Pl}$, where $E$ is some typical energy scale of any process under consideration. This procedure leads to a well-defined effective field theory of quantum gravity whose effects can be computed systematically to any desired order in the expansion in powers of $E/M_{\rm Pl}$
and can be applied in the limit $E\ll M_{\rm Pl}$.
Given the broad applicability of the methods of effective field theory, it stands natural that any ultraviolet consistent quantum theory of gravity should be expected to reproduce the effective field theory in the infrared limit.

In a rather straightforward way, the simplest reason why Einstein-Hilbert gravity is not perturbatively renormalizable is that the coupling -- Newton's constant $G_N$ -- is dimensionful and has negative mass dimension, implying that the theory is not power-counting renormalizable in the standard sense.
The observation that there are on-shell divergences that cannot be renormalized
at $1$- and $2$-loops starting from an expansion of the Gaussian theory simply fits the original premise of the lack of perturbative renormalizability. The dimensionful nature of Newton's constant also suggests
implicitly the effective status of the metric's quantum theory in the infrared.

A way around the lack of perturbative renormalizability is to study whether there are chances that a quantum theory of gravity might instead be \emph{nonperturbatively} renormalizable. This idea has originally been framed by Weinberg \cite{Weinberg:1980gg},
even before the perturbative nonrenormalizability was completely established.
The idea, which now goes under the name of \emph{asymptotic safety conjecture},
is funded on the premise that a metric theory of quantum gravity is asymptotically safe, rather than asymptotically free like certain gauge theories.
The properties of an asymptotically safe theory are best described in terms of the renormalization group: it is assumed that the renormalization group flow of the model under consideration admits a scale invariant solution which has a finite number of relevant deformations in the ultraviolet. In practical terms,
the fixed point is a zero of the couplings' beta functions and the number of relevant directions can be established by studying the stability matrix of the flow. The scale invariant fixed point plays the role of an ultraviolet completion
of the quantum theory that does not, in principle, require additional degrees of freedom
besides the metric and the pre-existing matter and gauge fields coupled to it.\footnote{Asymptotic safety is not a stranger to gauge theories: some specific classes of gauge-Yukawa theories in four dimensions have been shown using perturbation theory in the large color and flavor numbers (Veneziano limit) to be asymptotically safe, i.e., interacting in the UV limit~\cite{Litim:2014uca}.}
We discuss Weinberg's original formulation of asymptotic safety in Sect.~\ref{sect:asymptotic-safety}.

There are compelling reasons to believe the asymptotic safety conjecture, which are, somehow surprisingly, based on perturbative arguments. The first and most important for us is related to the fact that in two dimensions Newton's constant is dimensionless, which hints at the fact that the quantum theory could be perturbatively renormalizable. In two dimensions the analysis is complicated by the fact that the Einstein-Hilbert action is a topological invariant and the only propagating mode of the metric is the conformal one, but these complications can be circumvented. The result is that the renormalization of Newton's constant gives a negative beta function, $\beta_{G_N} = -b G_N^2$ and $b>0$, implying that the theory is asymptotically free. Now the crucial argument is that if we \emph{could} continue the theory and its renormalization group flow above $d=2$, we should replace $G_N \to \tilde{G}_N \mu^{\frac{d-2}{2}}$, where $\mu$ is the scale of the running and $\tilde{G}_N$ is Newton's constant measured in units of $\mu$. Under this assumption it is straightforward to see that the beta function of $\tilde{G}_N$ has an ultraviolet fixed point $\tilde{G}_N \sim \frac{\varepsilon}{b}$ where $d=2+\varepsilon$. The logic can be made almost flawless for $0<\varepsilon \ll 1$ and, extrapolating it to $\varepsilon=2$,
leads naturally to the conclusion that quantum gravity should be asymptotically safe in $d=4$ as argued by Weinberg himself \cite{Weinberg:1980gg}. This observation has spawned several important papers on two dimensional gravity and the $d=2+\varepsilon$ expansion, including the results by Jack \& Jones \cite{Jack:1990ey} and by Kawai \& Ninomiya \cite{Kawai:1989yh}, which can be regarded as seminal works on their own.
A partly historical account of the early results on $d=2+\varepsilon$ gravity is given in Sects.~\ref{sect:2pluseps} and \ref{sec:JJandAK} of this chapter.

A second compelling reason comes from the large-$N$ analysis of Tomboulis \cite{Tomboulis:1977jk} and Smolin \cite{Smolin:1981rm}, which shows that the gravitational fixed point of quantum gravity
can be found in general dimension $d$ -- including four dimensions as special case --
if one chooses to adopt an expansion
in the large number of matter ``flavors'' $N$ (here the term matter includes all nongravitational fields) for which $\tilde{G}_N \sim 1/N$ and further corrections could be computed as inverse powers of $N$.
The large-$N$ analysis gives the precious insight that the gravitational fixed point cannot be controlled only by the scalar curvature interaction through the Einstein-Hilbert action, but higher derivative operators that are nonzero
on-shell (under the equations of motion of general relativity in presence of matter) must be taken into account. In $d=4$, the crucial operator to consider is the square of Weyl's curvature, whose interaction is renormalized by inverse powers of $N$, making the gravitational fixed point a higher derivative gravitational nonperturbative action in the large-$N$ limit. The large-$N$ expansion is summarized in Sect.~\ref{sect:largeN} following the seminal work of Smolin \cite{Smolin:1981rm}

At this stage a clarification is in order. It is known since the work of Stelle \cite{Stelle:1976gc} that higher derivative gravity is actually perturbatively renormalizable in $d=4$. The higher derivative interactions that appear at quadratic order in the curvatures are, in fact, dimensionless and it is possible to construct a sound loop-perturbative expansion using powers of their couplings, which appear as \emph{inverse} powers multiplying the squared curvatures. In this case, the convergence of the expansion is improved by the fact that the propagators of the metric fluctuations scale as $1/p^4$ in momentum space, as opposed to $1/p^2$ of the standard actions with two derivatives. In perturbative higher derivative gravity, however, the Einstein-Hilbert term is just a relevant mass-like deformation of the corresponding Gaussian fixed point parametrized by the Planck's mass square. This is in contrast to the fundamental role that the Einstein-Hilbert term has in the construction of the asymptotically safe fixed point,
in fact the two solutions are believed to coexist \cite{Codello:2006in,Groh:2011vn}.
Furthermore, the $1/p^4$ leads to potential problems with unitarity, which have been addressed in the past. Higher derivative gravity might then be asymptotically free and, possibly, unitary, at least if some conditions are met, so its perturbative expansion remains a captivating alternative to the solution provided by asymptotic safety.
A brief report of some important results on higher derivative gravity and the modern approaches to circumvent its difficulties is given in Sect.~\ref{sec:higherDvs} of this chapter, but for a more detailed discussion of the renormalization we refer to Ohta's contribution to this book \cite{Ohta:2022zqs}.

Coming back to asymptotic safety,
it should be clear by now that the asymptotically safe fixed point
is well-established both in the $d=2+\varepsilon$ and in the large-$N$ expansions. Unfortunately both expansions do not apply straightforwardly
at finite $N$ and $\varepsilon = 2$,
which are the physically interesting real-world limits,
but rather require extrapolations. It could be possible, in principle,
to compute several orders of the perturbative expansions and apply
some resummation scheme, but this would require
a considerable amount of work because the computation of diagrams with several loops, graviton lines and derivative interactions is needed.
Even setting aside the difficulty of the computation
of subleading orders in the perturbative expansions,
there is the additional deterrent that both the perturbative and the large-$N$ expansions are 
most likely of asymptotic nature,
which would also require a solution to the rather hard problem of reconstructing the full result from the perturbative series (as proposed, for example, in the resurgent transseries program).

The way around to the limitations imposed by perturbation theory
that has been adopted in the literature is
to use a method for the computation of the renormalization group that is natively
nonperturbative. The first nonperturbative computation of the renormalization group flow of Einstein-Hilbert gravity is due to Reuter \cite{Reuter:1996cp},
who used the functional renormalization group flow of an effective \emph{average} action, which now goes under the name of Wetterich's equation \cite{Wetterich:1992yh}, and the background field method in combination with covariant computational techniques based on the heat kernel. The application of the effective average action clearly shows an asymptotically safe fixed point under some approximations \cite{Souma:1999at}, that we are going to discuss momentarily. Ever since the result by Reuter, which now goes under the name of ``Quantum Einstein Gravity'', most of the asymptotic safety literature has been based on the Wetterich equation, improving over the original computation in several ways, which are too many to be accounted here, but recent discussion on several aspects of the program can be found in~\cite{Bonanno:2020bil}. This side of the literature is also covered at length by other authors in other chapters of this book.

The use of functional renormalization group method comes at a price. The main problem is that there is no perturbative coupling/parameter which can be used to estimate the size of any involved approximation. In the Wetterich's approach there is no ``bare'' action, but rather one works directly with the effective average action that interpolates the effective action (the Legendre transform of the generator of connected diagrams) in the infrared, therefore it is necessary to \emph{truncate} the space of interactions of the average action
for the computations to be practically doable. Ultraviolet and infrared modes are distinguished by means of a cutoff function, which potentially introduces dependence on the computational scheme in its arbitrariness. Finally, the application to metric's gravity requires the use of the background field method which splits the metric, e.g., linearly as $g_{\mu\nu}=\overline{g}_{\mu\nu}+ h_{\mu\nu}$, in a background $\overline{g}_{\mu\nu}$ over which fluctuations $h_{\mu\nu}$ are integrated:
while the status of the average action as a function of an \emph{average field}
has been clarified in condensed matter applications in the past \cite{Wetterich:1989xg}, it is not straightforward that the same property holds for average fluctuations $\langle h_{\mu\nu} \rangle$ over a background metric. In fact, results based on the effective average action tends to depend on gauge-fixing and other scheme-related parameters, although
the evidence of the asymptotically safe fixed point is by believed by most to be conclusive.

For these reasons, at several points in time some authors advocated for a ``parallel'' reinassance of the perturbative approaches to metric's gravity, to be discussed in tandem with the nonperturbative ones. The reason being that most of the ``problems'' of the functional approaches can be discussed and solved relatively easily in the framework of perturbation theory, where it is straightforward to show that gauge and parametric dependencies cancel on-shell. Notably, this revival of perturbation theory included the works of Niedermaier \cite{Niedermaier:2009zz, Niedermaier:2010zz}, of Falls \cite{Falls:2015qga, Falls:2017cze, Falls:2018olk}, and, more recently,
of some of us using $d=2+\varepsilon$ \cite{Martini:2021slj, Martini:2021lcx}.
The application of perturbative methods highlights the importance of going on-shell when compiling the scaling properties of the effective action in the ultraviolet. It also shows that all unwanted parametric dependencies combine in the renormalization of the source term for the operator corresponding to the equations of motion, which, in turn, implies that the natural argument of the effective action is not simply the sum of the background metric and the expectation value of the fluctuations, $\langle g_{\mu\nu}\rangle \neq \overline{g}_{\mu\nu}+ \langle h_{\mu\nu}\rangle$, but, rather, a more complicate functional operator.\footnote{%
This is to be expected on the basis of the Vilkovisky-de Witt
approach to the construction of an effective action that is covariant under reparametrizations of the fields.
}
The perturbative approach also highlights potential caveats in the continuation of the dimensionality of spacetime $d$, both close to $d=2$, or up and above $d=4$. These have to do with the stability of the conformal mode, among other things,
and might potentially induce new lines of research in the parallel approaches based on functional renormalization. For example, a general prediction
of the revived perturbative approach is that there must be an upper critical dimension $d_{\rm cr}$ to the existence of the gravitational fixed point, which is not easy to see using the functional methods. The estimate from perturbation
theory is that $d_{\rm cr} \approx 7.5$, but there can be sizeable corrections beyond the leading order in the expansion, so it is not straightforward to determine that $d_{\rm cr} >4$, which is a necessary requirement for asymptotic safety in the physical case for obvious reasons. An account of the most recent works on $d=2+\varepsilon$ in given in Sect.~\ref{sec:2dRevised}, while
Sect.~\ref{sec:conclusions} offers an attempt to a conclusion.

The other sections of this contribution are meant to give a more in-depth discussion of the topics touched in this introduction as referenced above. Several of the computations that are presented in this chapter benefit from
a better knowledge of the background field method and of the heat kernel, which are presented in
\ref{app:bkgFieldMeth}
and in
\ref{app:covComputationsHK}
respectively.

%%%%%%%%%%%%%%%%%%%%%%%%%%%%%%%%%%%%%%%%%%%%%%%%%%%%%%%%%%%%%
%%%%%%%%%%%%%%%%%%%%%%%%%%%%%%%%%%%%%%%%%%%%%%%%%%%%%%%%%%%%%
\section{The failure of perturbative renormalizability}
\label{sec:falureRen}
%%%%%%%%%%%%%%%%%%%%%%%%%%%%%%%%%%%%%%%%%%%%%%%%%%%%%%%%%%%%%
%%%%%%%%%%%%%%%%%%%%%%%%%%%%%%%%%%%%%%%%%%%%%%%%%%%%%%%%%%%%%
{\it

The perturbative expansion of metric gravity starting from the bare Einstein-Hilbert action fails at $1$- or $2$-loops, depending on the presence of matter or the cosmological constant. The failure of perturbation theory is caused by divergences that cannot be reabsorbed in the renormalization of the original bare action,
but in new operators that include higher powers in the curvatures.
}

\smallskip

In this section we review the first attempts of renormalizing general relativity. We start by reviewing the one-loop results following the seminal works of 't Hooft \& Veltman \cite{tHooft:1974toh} for pure gravity and for gravity coupled to scalar fields, and of Christensen \& Duff \cite{Christensen:1979iy} for pure gravity with cosmological constant. Further historically relevant one-loop computations confirming and completing the results presented here can be found in \cite{Barvinsky:1993zg, Deser:1974cy, Deser:1974zzd}. We close this section by presenting the result found by Goroff \& Sagnotti \cite{Goroff:1985sz, Goroff:1985th} and van de Ven \cite{vandeVen:1991gw}, i.e, the two-loop counterterm from which we learn that gravity is not perturbatively renormalizable in four dimensions. Some of the original works were performed in Lorentzian signature however, for an easier comparison and a smoother exposition, we present all the results in their Euclidean signature counterpart.

The computation performed by 't Hooft \& Veltman is heavily based on a general formula for one-loop counterterm in flat space.
Given a general Lagrangian for complex scalar fields $\varphi_i$ on a $d$-dimensional flat space without boundary
\begin{align}
\label{eq:flatScalarTheory}
\mathcal{L} = \varphi^*_i\partial_\mu\partial^\mu\varphi_i 
 +2\varphi^*_i N^\mu_{ij}\partial_\mu\varphi_j + \varphi^*_i M_{ij}\varphi_j\,,
\end{align}
one can obtain the one-loop counterterm
\begin{align}
\label{eq:flatCounterterm}
\Delta\mathcal{L} = \frac{1}{(4\pi)^2\varepsilon}{\rm Tr}\left(X^2+\frac{1}{6}Y_{\mu\nu}Y^{\mu\nu}\right)\,,
\end{align}
where we compactly denoted
\begin{align}
\begin{split}
 X &= M-N^\mu N_\mu-\partial_\mu N^\mu\,,\\
 Y_{\mu\nu} &= \partial_\mu N_\nu-\partial_\nu N_\mu + N_\mu N_\nu - N_\nu N_\mu\,,\\
 \varepsilon &=d-4\,.
\end{split}
\end{align}
For a modern introduction to the necessary computational techniques to obtain \eqref{eq:flatCounterterm} the reader is referred to \ref{app:bkgFieldMeth}
and \ref{app:covComputationsHK} and references therein.
Using this result it is possible to simplify the computation of one-loop counterterms for matter fields coupled to gravity by drastically reducing the amount of graphs that have to be explicitly computed.

The generalization of the theory \eqref{eq:flatScalarTheory} to curved space reads
\begin{align}
\label{eq:curvedScalarTheory}
\mathcal{L}_s = \sqrt{g}\left(-g^{\mu\nu}\partial_\mu\varphi^*\partial_\nu\varphi 
	+2\varphi^* N^\mu\partial_\mu\varphi + \varphi^* M\varphi\right)\,.
\end{align}
One can approach the renormalization of such a theory by listing all the possible counterterms with the appropriate mass dimensions. 
Notice that when renormalizing a four dimensional theory, counterterms due to gravity include operators that are quadratic in curvature invariants. However, the term $R_{\mu\nu\rho\sigma}R^{\mu\nu\rho\sigma}$ can be ignored since in $d=4$ one has that the quantity
\begin{align}
\label{eq:eulerDensity}
\sqrt{g}E = \sqrt{g}\left(R_{\mu\nu\rho\sigma}R^{\mu\nu\rho\sigma}-4R_{\mu\nu}R^{\mu\nu}+R^2\right)\,
\end{align}
is a total derivative and its integral is the topological Euler character. Hence, one can solve the squared Riemann tensor in favor of the Ricci tensor and the Ricci scalar.
The coefficients of the counterterms that do not depend on the curvature can be fixed by direct comparison with the flat space case \eqref{eq:flatCounterterm}. Then, one can consider the special case of a conformally flat metric
\begin{align}
g_{\mu\nu}=\delta_{\mu\nu}F\,, \qquad F=1-f
\end{align}
and $f$ is an arbitrary function of spacetime. In this setup, the theory can be rewritten in the form \eqref{eq:flatScalarTheory} with the substitution
\begin{align}
\begin{split}
M\rightarrow FM\,, \qquad
N^\mu \rightarrow F N^\mu +\frac{1}{2}F^{-1}\partial^\mu F\,.
\end{split}
\end{align}
This fixes the counterterm Lagrangian up to a term of the form
\begin{align}
\Delta \mathcal{L}_a = \sqrt{g} a \left(R_{\mu\nu}R^{\mu\nu}-\frac{1}{3}R^2\right)\,.
\end{align}
The last coefficient $a$ can be determined upon computing the one-loop self energy for the gravitational fluctuations due to the scalar fields. In dimensional regularization only one Feynman diagram contributes to $a$.
The final result for the counterterm Lagrangian for the theory \eqref{eq:curvedScalarTheory} is
\begin{multline}
\label{eq:counterScalarCS}
\Delta \mathcal{L}_s = \frac{\sqrt{g}}{8\pi^2\varepsilon}{\rm Tr}\left\{\frac{1}{12}Y_{\mu\nu}Y^{\mu\nu}
 +\frac{1}{2}\left(M-N_\mu N^\nu-\nabla_\mu N^\nu-\frac{1}{6}R\right)^2\right. \\
+\left.\frac{1}{60}\left(R_{\mu\nu}R^{\mu\nu}-\frac{1}{3}R^2\right)\right\}\,.
\end{multline}

One can use a similar approach to renormalize quantum fluctuations of gravity. We could consider the Lagrangian
\begin{align}
\mathcal{L}_g = \sqrt{g}\left(-R-\frac{1}{2}g^{\mu\nu}\partial_\mu\varphi \partial_{\nu}\varphi\right)
\end{align}
and split the quantum fields in background plus fluctuations
\begin{align}
g_{\mu\nu} = \bar{g}_{\mu\nu} + h_{\mu\nu}\,,\qquad\varphi= \bar{\varphi} + \phi\,.
\end{align}
Due to the gauge symmetry of the gravitational content, it is necessary to include gauge fixing and ghost Lagrangians that depend on the background (we denote with $\bar{\nabla}$ the background covariant derivative). The gauge-fixing is
\begin{align}
\mathcal{L}_{gf} = -\frac{1}{2}\sqrt{\bar{g}}\;\bar{g}^{\mu\nu}F_\mu F_\nu\,,
\qquad
F_\mu = \bar{\nabla}_{\alpha}h^\alpha_\mu - \frac{1}{2}\bar{\nabla}_{\mu}h^\alpha_\alpha - \phi\partial_\mu\bar{\varphi}\,,
\end{align}
and the corresponding ghost Lagrangian is
\begin{align}
\mathcal{L}_{gh} = \sqrt{\bar{g}}\;\bar{C}_\mu\left(\delta^\mu_\nu\bar{\nabla}^2-\bar{g}^{\mu\alpha}\bar{R}_{\alpha\nu}-\partial^\mu\bar{\varphi}\partial_\nu\bar{\varphi}\right)C^\nu\,.
\end{align}

Although it requires cumbersome manipulations, the theory $\mathcal{L} = \mathcal{L}_g + \mathcal{L}_{gf} + \mathcal{L}_{gh}$, up to second order in the fluctuations, can be recast in the same form as \eqref{eq:curvedScalarTheory} and the corresponding counterterms are found to be
\begin{multline}
\label{eq:counterGravMatter1Loop}
\Delta\mathcal{L} = \frac{\sqrt{\bar{g}}}{8\pi^2\varepsilon}\left\{\frac{9}{720}\bar{R}^2
 +\frac{43}{120}\bar{R}_{\mu\nu}\bar{R}^{\mu\nu}
 +\frac{1}{2}(\bar{g}^{\mu\nu} \partial_\mu\bar{\varphi} \partial_\nu\bar{\varphi})^2\right.\\
\left.-\frac{1}{12}\bar{R}\bar{g}^{\mu\nu}\partial_\mu\bar{\varphi} \partial_\nu\bar{\varphi}
 +2(\bar{\nabla}_\mu\bar{\nabla}^\mu\bar{\varphi})^2\right\}\,.
\end{multline}
From this result, one can infer the one-loop counterterm for pure gravity by sending $\bar{\varphi}\to 0$ and subtracting \eqref{eq:counterScalarCS} at $M=N=0$. As a result, we have
\begin{align}
\Delta\mathcal{L}_{pg} = \frac{\sqrt{\bar{g}}}{8\pi^2\varepsilon}\left(\frac{1}{120}\bar{R}^2
 +\frac{7}{20}\bar{R}_{\mu\nu}\bar{R}^{\mu\nu}\right)\,.
\end{align}

A few remarks are in order. First, we notice how employing Einstein's equations in vacuum, the counterterm Lagrangian for pure gravity vanishes and the theory is one-loop finite. We could have guessed this without any actual computation since, for dimensional reasons, the only counterterm that could survive on-shell is given by the square of the Riemann tensor but, as we said earlier, this can be eliminated thanks to the topological density \eqref{eq:eulerDensity}.

However, if we now switch on the scalar fields, we can see that the background equations of motion for the gravity matter system reduce to
\begin{align}
\begin{split}
 \bar{\nabla}_\mu\bar{\nabla}^\mu\bar{\varphi}=0\,,\qquad
 \bar{R}_{\mu\nu} = -\frac{1}{2}\bar{\nabla}_\mu \bar{\varphi}\bar{\nabla}_\nu\bar{\varphi}\,.
\end{split}
\end{align}
Inserting this into \eqref{eq:counterGravMatter1Loop} we find that on-shell
\begin{align}
\Delta\mathcal{L} = \frac{\sqrt{\bar{g}}}{8\pi^2\varepsilon}\frac{203}{80}\bar{R}^2\,,
\end{align}
hence, the gravity matter system is not one-loop renormalizable.
This last point suggests the fact that pure gravity is one-loop finite might be a coincidence due to the form of Euler character in four dimensions.

It might be interesting to check how the renormalization of quantum gravity is affected by the inclusion of a cosmological constant. The theory is described by the action
\begin{align}
\label{eq:actionEinstHilCosmCons}
S_{\Lambda}[g] = -\frac{1}{16\pi G_N}\int{\rm d}^4 x\sqrt{g}\;\left(
R
-2\Lambda\right)\,,
\end{align}
and, at the classical level, is governed by the equations of motion
\begin{align}
\label{eq:eomCosmological}
R_{\mu\nu}-\frac{1}{2}R g_{\mu\nu}+\Lambda g_{\mu\nu}=0\,.
\end{align}
Taking the trace of \eqref{eq:eomCosmological} one finds, in dimension $d=4$, $R_{\mu\nu}=\Lambda g_{\mu\nu}$ and $R = 4\Lambda$. This gives the following simplification of the Euler density
\begin{align}
E = R_{\mu\nu\rho\sigma}R^{\mu\nu\rho\sigma}\,.
\end{align}
Then, the one-loop computation is a straightforward generalization of the case without cosmological constant and can be performed using heat kernel methods in the background field method. The counterterm Lagrangian is given by
\begin{align}
\label{eq:chrisDuffCountertermCosmo}
\Delta \mathcal{L}_{\Lambda} = \frac{\sqrt{\bar{g}}}{180(4\pi)^2\varepsilon}\,\left(\bar{R}_{\mu\nu\rho\sigma}\bar{R}^{\mu\nu\rho\sigma}-2088\Lambda^2\right)\,.
\end{align}
The term squared in the Riemann tensor is topological. One could therefore decide to include a term of the form $\alpha \int E$ in the original action without changing the equations of motion, and absorb the first term of \eqref{eq:chrisDuffCountertermCosmo} in the running of $\alpha$. The second term contributes to the running of the cosmological constant which will exhibit a quadratic beta function. Again, pure gravity in four dimensions shows no formal issue at one-loop level thanks to the form of the topological Euler density. It is crucial, however, to check how the theory behaves beyond the one-loop approximation.
An explicit calculation shows that, at two-loops, the on-shell divergence for the Einstein-Hilbert theory is
\begin{align}
\label{eq:gorffSagnotti}
\Gamma_{\infty} = \frac{1}{(4\pi)^4\varepsilon}\frac{209}{2880}
 \int {\rm d}^4\sqrt{\bar{g}}\;{\bar{R}^{\alpha\beta}}_{\quad\gamma\delta}
 {\bar{R}^{\gamma\delta}}_{\quad\zeta\xi}{\bar{R}^{\zeta\xi}}_{\quad\alpha\beta}\,.
\end{align}
It is now clear that quantum gravity in dimension $d=4$ is not perturbatively renormalizable.

%%%%%%%%%%%%%%%%%%%%%%%%%%%%%%%%%%%%%%%%%%%%%%%%%%%%%%%%%%%%%
%%%%%%%%%%%%%%%%%%%%%%%%%%%%%%%%%%%%%%%%%%%%%%%%%%%%%%%%%%%%%
\section{Higher derivative gravity}
\label{sec:higherDvs}
%%%%%%%%%%%%%%%%%%%%%%%%%%%%%%%%%%%%%%%%%%%%%%%%%%%%%%%%%%%%%
%%%%%%%%%%%%%%%%%%%%%%%%%%%%%%%%%%%%%%%%%%%%%%%%%%%%%%%%%%%%%

{\it
One possible solution to the problem outlined in Sect.~\ref{sec:falureRen} is to reformulate the perturbative expansion in terms of a higher derivative action, which is quadratic in the curvatures, instead of the Einstein-Hilbert action, which would then be a relevant deformation of the higher derivative action.
}

\smallskip

The advantage of this choice is that the bare propagator has better convergence properties due to the higher powers of momenta flowing in the loops. However, the unitarity of the theory is not guaranteed any more and requires special considerations. In this section we review some of the main aspects of this approach, making contact with the problem of unitarity.
The fundamental references are the original work from Stelle \cite{Stelle:1976gc}, Refs.~\cite{Avramidi:1985ki, Fradkin:1981iu} where some errors in the computation of the renormalization group of earlier publications are corrected, but also the more recent Refs.~\cite{deBerredoPeixoto:2003pj, deBerredoPeixoto:2004if}. Higher derivative gravity can be asymptotically free or trivial depending on one coupling,
the phase in which it is not asymptotically free is argued to be the physical one and investigated in \cite{Salvio:2018crh, Salvio:2017qkx} aiming at the ultraviolet completion.\footnote{%
Neglecting the problem of unitarity, because \emph{shut up and compute!}
}
The Lee-Wick approach to the problem of unitarity is explained in \cite{Anselmi:2018ibi}, while the discussion of higher derivative gravity as a PT-symmetric theory appears in \cite{Mannheim:2020ryw, Mannheim:2021oat}.
% }

There are two formulations of higher derivative gravity that is important to introduce for our purposes: the conformal version and the nonconformal one. 
The conformal action is
\begin{align}
\label{eq:confGravAction4d}
S_{CG}[g] = -2\alpha \int {\rm d}^d x\sqrt{g} \left[R_{\mu\nu}R^{\mu\nu}-\frac{1}{3}R^2\right]\,,
\end{align}
which is on-shell equivalent to the integrated square of the Weyl tensor.
Conformal gravity is invariant under Weyl transformations of the metric; it emerges, for example, as the one-loop effective action of massless Dirac fermions that are minimally coupled to curved space through vielbeins and spin-connection.

The nonconformal action, also known as Stelle's action, is
\begin{align}
\label{eq:hdAction}
S_{hd}[g] = -\int{\rm d}^4x\sqrt{g}\left[\frac{1}{2\lambda}C^2-\frac{1}{\rho}E+\frac{1}{\xi}R^2+
 \tau\nabla^2R-\frac{1}{G}(R-2\Lambda)\right]\,,
\end{align}
where $E$ is the Euler density defined in \eqref{eq:eulerDensity}, $C^2=C_{\mu\nu\rho\sigma}C^{\mu\nu\rho\sigma}$ is the square of the Weyl tensor and $G=16\pi G_N$. The higher derivatives operators control the perturbative series and would be enough to inspect for asymptotic freedom. The term linear in $R$ is included as a relevant deformation and to allow for the classical field equations to be solved by the flat space metric. The equations are
\begin{align}
\label{eq:hdEOM}
\begin{split}
&\frac{1}{G}\left(R^{\mu\nu}-\frac{1}{2}Rg^{\mu\nu}+g^{\mu\nu}\Lambda\right)
 +\frac{1}{\lambda}\left[2\left(\frac{1}{3}-\frac{\lambda}{\xi}\right)R\left(R^{\mu\nu}
 -\frac{1}{4}g^{\mu\nu}R\right)+\frac{1}{2}g^{\mu\nu}R_{\alpha\beta}R^{\alpha\beta}\right.\\
&\left. -2R^{\mu\alpha\nu\beta}R_{\alpha\beta}
 +\left(\frac{1}{3}+2\frac{\lambda}{\xi}\right)\nabla^\mu\nabla^\nu R-\nabla^2R^{\mu\nu}
 +\left(\frac{1}{6}-2\frac{\lambda}{\xi}\right)g^{\mu\nu}\nabla^2 R\right]=0\,.
\end{split}
\end{align}

The covariant quantization of the theory in the background field approach requires an appropriate gauge fixing. From a kinematical point of view, one can choose the gauge fixing condition $F_\mu [h]$ in the standard manner, however from a dynamical point of view the gauge fixing action will have to contain higher order derivatives to suppress the gauge modes appropriately. These extra derivatives play the role of a (background) bilinear form $H$ in the gauge fixing action so we can write a gauge fixed path integral in the form
\begin{align}
Z[g] = \det H^{-\frac{1}{2}}\int\mathcal{D}h\mathcal{D}\bar{\eta}\mathcal{D}\eta\;
 \exp\left\{-S_{hd}[\bar{g}+h]-\frac{1}{2}F_{\mu}H^{\mu\nu}F_{\nu}
 -\bar{\eta}_\mu H^{\mu\lambda}\Delta^{gh}_{\lambda\nu}\eta^\nu\right\}\,,
\end{align}
where $\eta$ and $\bar{\eta}$ are the Faddeev-Popov ghost and antighost respectively and $\Delta^{gh}$ is found in the usual way starting from the gauge condition $F[h]$. The one-loop effective action is found to be
\begin{align}
\begin{split}
\Gamma_{hd} = \frac{1}{2}\Tr\log\Delta_h &-\Tr\log\Delta^{gh}-\frac{1}{2}\Tr\log H\,,\\
\Delta_h^{\mu\nu\alpha\beta} =& \left.
 \frac{\delta^2S_{gh}}{\delta g_{\mu\nu}\delta g_{\alpha\beta}}\right|_{\bar{g}}\,.
\end{split}
\end{align}
An explicit calculation with dimensional regularization in $d=4-\varepsilon$ dimensions leads to a counterterm action that includes the same operators as above
\begin{align}
\label{eq:hdCounterterm}
\Gamma_{ct} = \frac{\mu^\varepsilon}{16\pi^2\varepsilon}\int{\rm d}^4x\sqrt{\bar{g}}&\left[
 \beta_1 \bar{C}^2 - \beta_2 \bar{E} + \beta_3 \bar{R}^2 +\beta_4 \bar{R} + \beta_5\right]\,,
\end{align}
with the coefficients
\begin{align}
\begin{split}
\beta_1 = \frac{133}{20}\,,\qquad\beta_2 =& \frac{196}{45}\,, \qquad
 \beta_3 = 10\frac{\lambda^2}{\xi^2} - 5\frac{\lambda}{\xi} + \frac{5}{36}\,,\\
\beta_4=\frac{\lambda}{G}\left(\frac{\xi}{12\lambda}-\frac{13}{6}-10\frac{\lambda}{\xi}\right)\,,\quad
 &\beta_5=\frac{1}{G^2}\left(\frac{5}{2}\lambda^2+\frac{\xi^2}{72}\right) 
 + \frac{\Lambda \lambda}{G}(\frac{56}{3}-\frac{2\xi}{9\lambda})\,.
\end{split}
\end{align}
Notice that the coupling $\tau$ drops out of the counterterm action
as expected from a total derivative.

An important question concerns the physical content of the renormalization flow generated by \eqref{eq:hdCounterterm}. In the higher derivative sector, we can look for a ``gauge invariant'' combination of the couplings (which have beta functions with universal coefficients independent on gauge fixing parameters).
A natural choice is
\begin{align}
\theta = \frac{\lambda}{\rho} \qquad \omega = -3\frac{\lambda}{\xi}\,.
\end{align}
In $d=4$ one finds two fixed points, one UV stable and one UV unstable, both for negative of $\omega$. The UV stable fixed point corresponds to asymptotic freedom of the theory. A negative solution for $\omega$ is often disregarded on the ground that it leads to an unstable static gravitational potential\footnote{Note that the sign of $\omega$ is opposite to the sign of $\xi$ since $\lambda$ is required to be greater than zero requiring positivity of the theory.}. 

What is, then, the physical picture corresponding to $\omega>0$? One could study this scenario in the context of ``agravity'' \cite{Salvio:2017qkx}, i.e, only including dimensionless operators in the theory truncation (therefore dropping the Einstein-Hilbert sector) and coupling the theory to a matter sector (again, truncated to the dimensionless operators). The UV limit of such theory can reach a conformal theory of gravity provided that $\xi$ diverges and all the other couplings remain asymptotically free. This happens because the two degrees of freedom of the conformal mode\footnote{In higher derivative gravity, the conformal transformation of $R^2$ generates two scalar modes: one with the standard kinetic sign of a scalar field and one corresponding to the usual conformal factor with opposite sign in the kinetic term.} decouple at high energy and the limit $\xi\to\infty$ is consistent only when the coupling to scalar fields, $
\mathcal{L}_\phi = -\frac{f_{ab}}{2}\phi_a\phi_b R
$,
reaches the Weyl invariant value of $f_{ab}=-\frac{1}{6}\delta_{ab}$. Conversely, the Weyl invariant high energy theory is not consistent since the symmetry is anomalous and at lower energies, the conformal mode couples to those operators that break scale invariance through their $\beta$-functions. As a result there is an anomalous generation of finite values of $\xi$ along the RG flow. According to \cite{Salvio:2017qkx}, it is possible to exclude (at least perturbatively) that the coupling $\xi$ hits a Landau pole in the high energy limit, however non-perturbative computations would be needed to discern whether it hits a UV interacting fixed point or if $\xi$ still grows to infinity. However,
we must always recall that the renormalization group actually
\emph{only} flows from
UV to IR, and not the converse, so it would be appropriate for future developments of \cite{Salvio:2017qkx} to formulate the approach from a ``top-down'' perspective.

What about unitarity? Perturbatively, unitarity appears to be lost due to the propagation of extra ghosts fields associated with the higher derivatives propagator (kinematical ghosts are negative norm states of the S-matrix). If one tries to perform only analytical operations on the propagator defined by \eqref{eq:hdAction}, the only possibility to restore unitarity seems to come at the price of unbounded energies in the theory spectrum. This feature is common in higher derivatives field theories, however in the context of Lee-Wick theories it is possible to deal with these issues by performing a non-analytic continuation of the tree-level propagator. Such a procedure, that consists in deforming the integration domain of Feynman graphs in momentum space, turns the extra ghosts into fake degrees of freedom known as fakeons \cite{Anselmi:2018ibi}. Fakeons appear as purely virtual particles in the vertex expansion of correlators, but they do not appear in the physical spectrum. In this way one gets rid of ghosts and obtains a unitary theory without spoiling the boundedness of the energy spectrum. As consequence of this prescription to compute correlators in higher derivative theories, microcausality is lost, however it is possible to compute a time scale that bounds such a violation and macrocausality seems to be preserved.

A slightly different take on the topic of unitarity relies on the realization that knowing the propagator of a theory is insufficient to the aim of constructing the Hilbert space. A careful analysis of \eqref{eq:hdAction} reveals that the Hamiltonian is not Hermitian but rather PT-symmetric\footnote{The Hamiltonian in actually CPT-symmetric, but since charge conjugation is preserved independently it is also a PT-symmetric theory.} \cite{Mannheim:2020ryw, Mannheim:2021oat}. The choice of the vacuum of the theory is then what discerns between a scenario with unbounded negative energies from one that apparently is non-unitary. However the correct choice of internal product for the Hilbert space reveals that \eqref{eq:hdAction} is actually free of ghosts, in the sense that they do not contribute with negative norm to the spectrum of the theory. Unfortunately, a negative aspect of this analysis is that the limit from \eqref{eq:hdAction} to pure quadratic gravity is singular and one cannot get rid of the Einstein-Hilbert sector to compare to the situation presented, e.g, in conformal pure gravity.

%%%%%%%%%%%%%%%%%%%%%%%%%%%%%%%%%%%%%%%%%%%%%%%%%%%%%%%%%%%%%%
%%%%%%%%%%%%%%%%%%%%%%%%%%%%%%%%%%%%%%%%%%%%%%%%%%%%%%%%%%%%%%
\section{Large-$N$ expansion}
\label{sect:largeN}
%%%%%%%%%%%%%%%%%%%%%%%%%%%%%%%%%%%%%%%%%%%%%%%%%%%%%%%%%%%%%%
%%%%%%%%%%%%%%%%%%%%%%%%%%%%%%%%%%%%%%%%%%%%%%%%%%%%%%%%%%%%%%

{\it 
The limit of large number of components of field multiplets typically leads to strong simplifications in otherwise inaccessible computations. We recall here interesting results in such limit when multi-component matter fields are coupled to gravity.
}

\smallskip

This approach to a quantum theory of gravity with improved renormalization properties relies on coupling the geometry to matter degrees of freedom. Typically, the number of matter fields of the model appears parametrically in the renormalization group flow of the gravitational sector and the limit of large number of flavor species can be accessed with standard perturbative methods.
This approach was initiated by Tomboulis in \cite{Tomboulis:1977jk}, where the analytical properties of the gravitational propagator have been studied. In Ref.~\cite{Smolin:1981rm}, which we follow in this section, Smolin investigates the existence of an interacting ultraviolet fixed point. Further results on the large-$N$ limit of quantum gravity can be found in \cite{Percacci:2005wu, Codello:2011js}, in which alternative non-perturbative analysis have been carried on.
The following analysis focuses on the four dimensional case, however a similar study can be carried on in two dimensions. Some comments about the effect of matter couplings to the renormalization of gravity in $d=2+\varepsilon$ are at the end of this section and refer to part of the work done in \cite{Smolin:1981rm}. Further examples can be found in \cite{Martini:2018ska}, where the analysis focuses on scalar matter rather then fermionic degrees of freedom.

We start by considering the action for $N$ massless spinors coupled dynamically to the geometry (in Lorentzian signature)
\begin{align}
\label{eq:fermionGravlagrangian}
S =-\int{\rm d}^4x\sqrt{g}\left[\frac{1}{G}R
 -\sum_{i=1}^N \bar{\psi}_i\slashed{\nabla}\psi_i+\frac{1}{2}\Lambda\right]\,,
\end{align}
where $\psi_i$ is a Dirac spin-$1/2$ field and $R$ is the Ricci scalar for the metric $g_{\mu\nu}$. We can study the UV behavior of the theory introducing a cut-off $K$ and inspecting the divergences in the large-$N$ limit via an expansion in $\frac{1}{N}$ of the Feynman diagrams. It is useful to introduce rescaled couplings as
\begin{align}
\begin{split}
\frac{1}{G} = \frac{c N K^2}{2}\,,\qquad
\Lambda = N \lambda K^4\,,
\end{split}
\end{align}
and rewrite \eqref{eq:fermionGravlagrangian} as
\begin{align}
S =-\int{\rm d}^4x\sqrt{g}\left[\frac{1}{2}c N K^2 R
 -\sum_{i=1}^N \bar{\psi}_i\slashed{\nabla}\psi_i+\frac{1}{2}\lambda N K^4\right]\,.
\end{align}
A convenient expansion is in terms of dimensionless fluctuations $g_{\mu\nu} = \eta_{\mu\nu} + h_{\mu\nu}$, such that all the leading order divergences will appear in the graviton propagator, rather then in the graviton-matter vertex. The drawback of this choice is that the flat Minkowski  metric is not a solution of the equations of motion due to the presence of a cosmological constant rescaled by $K^4$. To ensure a consistent renormalization one would need to consider tadpole diagrams as well to renormalize the classical equations of motion. However in \cite{Smolin:1981rm} it was shown that in the limit of vanishing renormalized cosmological constant, at leading order in $\frac{1}{N}$, the tadpole diagrams due to fermionic loops cancel against tadpoles due to the insertion of the gravitational $1$-point function. We therefore simplify the analysis presented here and assume that we have performed this choice of scheme.

At this point the analysis can be carried out by computing the renormalized propagator and requiring that the limit $K\to\infty$ exists. Introducing the projector on the spin-$2$ states $P^{(2)}_{\mu\nu\alpha\beta}$, the bare propagator for the gravitational sector is
\begin{align}
D_{\mu\nu\alpha\beta}(p^2) = \frac{P^{(2)}_{\mu\nu\alpha\beta}}{cNK^2(p^2 + i\varepsilon)}+\dots
\end{align}
where the dots stand for gauge dependent parts propagating the lower spin modes and we will not consider them here.

The leading order divergences in the $\frac{1}{N}$ expansion are given by the insertion of fermionic loops, but no gravitational loop. The result for the one-loop renormalized propagator in four dimensions is then given in \cite{Smolin:1981rm}
\begin{align}
\label{eq:renormalizedPropLargeN}
\begin{split}
D^{1/N}_{\mu\nu\alpha\beta}(p^2) =& 
 \frac{P^{(2)}_{\mu\nu\alpha\beta}}{cNp^2K^2\left[1-\left(cNp^2K^2\right)^{-1}
 \left(iN F^{(2)} + \lambda NK^4\right)\right]}+\dots\,.\\
\end{split}
\end{align}

$F^{(2)}$ is the spin-$2$ part of the fermionic loop insertion in the Feynman diagrams needed to renormalize the graviton propagator and contains polynomial terms in he cut-off $K$ up to quartic order and a logarithmic term $\sim p^4\log(-p^2/K^2)$. Requiring that the $K$ dependence in the denominator disappears as in \eqref{eq:renormalizedPropLargeN} fixes the value of $c$ up to a term that vanishes in the limit $K\to\infty$
\begin{align}
\label{eq:cRG}
\frac{1}{c} = 32\pi^2\left(1 - \frac{M^2}{K^2}\right)\,.
\end{align}
Similarly, $\lambda$ is fixed by requiring that the $K^4$ term in $F^{(2)}$ is canceled by the (divergent) bare cosmological constant
\begin{align}
\label{eq:lambdaRG}
\lambda = \frac{9}{128\pi^2} + \frac{L^4}{K^4} 32 \pi^2\,.
\end{align}
From Eqs.~\eqref{eq:cRG} and \eqref{eq:lambdaRG}, we see that
the system $\{c, \lambda\}$ approaches a fixed point
\begin{align}
\{c^*, \lambda^*\} = \left\{\frac{1}{32\pi^2}, \frac{9}{128\pi^2}\right\}\,.
\end{align}
The rate at which the fixed point is approached is given by the combination $\sqrt{N}M$, which at low energies plays the role of an effective Planck mass, and by $L$. The requirement of vanishing bare cosmological constant is $L=0$.

We are now left with the logarithmic divergence coming from $F^{(2)}$. In order to cancel this term we need to include a higher derivative operator in the bare Lagrangian density
\begin{align}
\mathcal{L}_C = \frac{N}{\alpha\left(\frac{M}{K}\right)}C^2\,,
\end{align}
where $C^2$ is the square of the Weyl tensor and $\alpha$ is a new bare parameter. The renormalization of $\alpha$ is taken care by the counterterm \cite{Tomboulis:1977jk, Smolin:1981rm}
\begin{align}
\label{eq:alphaCounterterm}
\Delta\mathcal{L} = N\left[\frac{1}{480\pi^2}\log\left(\frac{K^2}{M^2}\right)+\frac{1}{\alpha}
		-\frac{353}{150}\right]C^2\,.
\end{align}
From the counterterm \eqref{eq:alphaCounterterm}, one can see that the new coupling is asymptotically free and the full system of couplings has the fixed point
\begin{align}
\label{eq:largeNFixPoint}
\{c^*, \lambda^*, \alpha^*\} = \left\{\frac{1}{32\pi^2}, \frac{9}{128\pi^2}, 0\right\}\,.
\end{align}

The computation of the renormalized propagator for the graviton in the new truncation leads to
\begin{align}
\label{eq:renormalizedPropLargeN2}
D^{1/N}_{\mu\nu\alpha\beta} = \frac{P^{(2)}_{\mu\nu\alpha\beta}}
 {Np^2 \left(M^2 - \frac{1}{\alpha p^2}-\frac{1}{480\pi^2}p^2\log\left(-\frac{p^2}{M^2}\right)\right)}+\dots\,.
\end{align}
The result \eqref{eq:renormalizedPropLargeN2} coincides with the one found using in dimensional regularization, where only the coupling $\alpha$ gets renormalized as shown in \cite{Tomboulis:1977jk}.

Before moving on it is worth mentioning that a similar analysis can be performed in dimension $d=2+\varepsilon$, as shown in \cite{Smolin:1981rm}. The analysis carries through in a similar fashion as shown here up to few modifications. Again, the inclusion of a higher derivative operator, proportional to $R_{\mu\nu}^2$ and parametrized by the coupling $\alpha$, is needed to cancel logarithmic divergences. The sector of Einstein-Hilbert couplings reaches a fixed point that can be understood as the continuation of \eqref{eq:largeNFixPoint} and reads
\begin{align}
\{c^*, \lambda^*\} = \left\{\frac{1}{36\pi}, \frac{9}{32\pi}\right\}\,.
\end{align}
Although, as we will see in the next section, there are many indications of a UV fixed point of order $O(\varepsilon)$ for Newton's constant in $d=2+\varepsilon$, the fixed point found in the large $N$ expansion is of order $O(1)$ and, thus, seem to be unrelated to the fixed point we will study in the next sections.

%%%%%%%%%%%%%%%%%%%%%%%%%%%%%%%%%%%%%%%%%%%%%%%%%%%%%%%%%%%%%
%%%%%%%%%%%%%%%%%%%%%%%%%%%%%%%%%%%%%%%%%%%%%%%%%%%%%%%%%%%%%
\section{Weinberg's Asymptotic Safety}
\label{sect:asymptotic-safety}
%%%%%%%%%%%%%%%%%%%%%%%%%%%%%%%%%%%%%%%%%%%%%%%%%%%%%%%%%%%%%
%%%%%%%%%%%%%%%%%%%%%%%%%%%%%%%%%%%%%%%%%%%%%%%%%%%%%%%%%%%%%

{\it
We present the general ideas at the basis of the Weinberg's asymptotic safety paradigm
which extends the scope of renormalizable quantum field theories.
The paradigm draws heavily from the interpretation of the renormalization group
that emerges from the theory of critical phenomena, which represents a significant departure from the purely pragmatical perspective that was applied to particle physics before Weinberg.
}

\smallskip

The underlying assumption of the programs outlined so far is that renormalizability has to take place close to the Gau{\ss}ian fixed point of a theory. If we then desire that the finite theory is ultraviolet complete, the Gau{\ss}ian fixed point must be ultraviolet and the theory asymptotically free.
This surely looks like the most natural assumption when looking for UV completeness within the perturbative regime, however asymptotic freedom is not the only possibility that may arise.
In \cite{Weinberg:1976xy, Weinberg:1980gg} Weinberg suggested that an alternative to asymptotic freedom could be a scenario in which the couplings of the theory reach scale invariance at infinite energy and physical observables remain finite. In general, such regime might be located outside the validity of perturbation theory, even though in the special case of asymptotic freedom it is captured by perturbative methods. This (generally nonperturbative) regime is referred to as asymptotic safety and is best phrased in terms of effective field theory methods (see \cite{Weinberg:2009bg} for a more recent account of the approach). Here we list the fundamental concepts and subtleties that become important when dealing with gravity following the original formulation by Weinberg \cite{Weinberg:1980gg}.

Let us consider an observable $r$ associated with some physical process. In general, $r$ will depend on some characteristic energy $E$ of the process under consideration, the set couplings of the underlying field theory $g_i$, and a set dimensionless variables $\zeta$. The expression for $r$ can be cast as
\begin{align}
r = \mu^D f\left(\frac{E}{\mu}, \bar{g}_i(\mu), \zeta\right)\,,
\end{align}
where $D$ is the total mass dimension of $r$, $\mu$ is the renormalization scale and $\bar{g}_i(\mu)=\mu^{d_i}g_i(\mu)$ are dimensionless coupling constants. Making use of the fact that the observable should not depend on $\mu$, we are free to set it to $\mu=E$ and obtain a simple scaling relation
\begin{align}
r = E^Df\left(1, \bar{g}_i(E), \zeta\right)\,.
\end{align}
It is clear from this last equation that, aside for a trivial scaling $E^D$, the behavior of $r$ is determined by that of the dimensionless couplings $\bar{g}(\mu)$ as $\mu\to \infty$.
We write
\begin{align}
\label{eq:rgGeneric}
\mu\frac{{\rm d}}{{\rm d}\mu}\bar{g}_i(\mu) = \beta_i\left(\bar{g}_i(\mu)\right)\,,
\end{align}
where, on the ground of dimensional arguments, the $\beta$ functions cannot depend explicitly on $\mu$ (couplings are expressed in units of $\mu$ itself). Note that if the couplings reach a fixed point
\begin{align}
\label{eq:asympSafetyGeneralCondition}
\bar{g}_i(\mu)\underset{\mu\to\infty}{\longrightarrow}\bar{g}_i^* \,,
\end{align}
then the observable $r$ has the simple scaling $E^D$ and the theory is free from unphysical divergences (the ratio between the observable and $E^D$ is a finite prediction). The condition for scale invariance \eqref{eq:asympSafetyGeneralCondition} is equivalent to the vanishing of the $\beta$-function
\begin{align}
\beta_i(\bar{g}_i^*) = 0\,.
\end{align}
Note that the existence of a fixed point for the dimensionless couplings is not sufficient to determine a scale invariant theory in the ultraviolet. The theory, in fact, needs to sit on a trajectory in the coupling space such that its renormalization group evolution hits the fixed point. The set of couplings that are dragged into the fixed point $\bar{g}_i^*$ when evolving towards the UV form the ultraviolet critical surface of fixed point.

A few comments are in order.
\begin{itemize}
\item The stress that we put on true physical observables is motivated by the fact that they do not depend on local field redefinitions. This requirement is equivalent to retain only essential couplings, i.e., to discard all those couplings $\gamma$ for which a variation of the Lagrangian $\mathcal{L}(\varphi_n, \partial\varphi_n, \dots)$ is proportional to the Euler-Lagrange equations of motion up to total derivative terms
\begin{align}
\label{eq:inessential}
\frac{\partial \mathcal{L}}{\partial \gamma} 
	\doteq \sum_n\left[\frac{\partial \mathcal{L}}{\partial\varphi_n}-
	\partial_\mu\frac{\partial\mathcal{L}}{\partial\left(\partial_\mu\varphi_n\right)}
	+\dots\right]F_n(\varphi_n, \partial\varphi_n, \dots)\,,
\end{align}
where $\doteq$ means that total derivatives have been discarded. A coupling $\gamma$ satisfying \eqref{eq:inessential} is called inessential (a simple example would be the wavefunction renormalization of a self-interacting scalar).
If inessential couplings are included the system is not guaranteed to hit the fixed point under the renormalization group flow, even though the system is physically equivalent. The case of gravity is somewhat peculiar because Planck's mass requires special considerations \cite{Percacci:2004yy}.
\item The dimensionality of the ultraviolet critical surface gives us the number of free parameters that must be fine tuned for the theory to reach scale invariance in the UV. It is therefore crucial that the critical surface be of finite dimension $\delta<\infty$, otherwise the theory loses its predictive power. $\delta$ can be calculated perturbatively by linearizing the system of $\beta$-functions around the UV fixed point and is given by the number of negative eigenvalues of the stability matrix 
\begin{align}
S_{ij} = \left.\frac{\partial \beta_i}{\partial\bar{g}_j}\right|_{\bar{g}=\bar{g}^*}
\end{align}
\item As in every approach to renormalization, a special care should be paid when ``following'' the flow towards the UV. As is clear from Wilson's approach to renormalization, flowing towards the ultraviolet is ill defined because it implies the reconstruction of an information that has been integrated out from a lower energy effective field theory. In the case of an asymptotically free theory, this problem is partially solved by the vanishing strength of interactions making the free theory stable against perturbations. In the case of asymptotic safety, the search for a non-perturbative fixed point is challenged by the fact that one would need to include all the possible (essential) operators compatible with the symmetries of the theory, so that all the information of the fundamental theory has been already reconstructed (up to symmetry restorations). In the present discussion we are silently bypassing this issue and assume that a systematic analysis of the theory space underlies the analysis.
\end{itemize}

%%%%%%%%%%%%%%%%%%%%%%%%%%%%%%%%%%%%%%%%%%%%%%%%%%%%%%%%%%%%%
%%%%%%%%%%%%%%%%%%%%%%%%%%%%%%%%%%%%%%%%%%%%%%%%%%%%%%%%%%%%%
\section{Gravity in $d=2+\varepsilon$}
\label{sect:2pluseps}
%%%%%%%%%%%%%%%%%%%%%%%%%%%%%%%%%%%%%%%%%%%%%%%%%%%%%%%%%%%%%
%%%%%%%%%%%%%%%%%%%%%%%%%%%%%%%%%%%%%%%%%%%%%%%%%%%%%%%%%%%%%

{\it
We present the first practical computation with which the possibility that gravity is asymptotically safe was established.
The results are based on the perturbative $\varepsilon$-expansion above $d=2$ dimensions.
}

\smallskip

There are two important things happening to the Einstein-Hilbert action in the limit $d=2$. The Newton's constant become dimensionless and
the Einstein-Hilbert term becomes topological, because it is proportional to the Euler character
\begin{align}
\chi = \int{\rm d}^2 x \sqrt{g}\;R\,.
\end{align}
The left hand side of Einstein's equations therefore vanishes identically (thanks to the contracted differential Bianchi identities).

In general $d$, Newton's coupling constant has dimensions $[G]=2-d$ and becomes dimensionless for $d=2$. Using dimensional regularization, one is therefore tempted to discard possible  countertems proportional to the Lagrangian on the ground that $\chi$ vanishes for $\varepsilon = d-2$ going to zero.\footnote{%
We implicitly assume that the topology of the spacetime is that of a non-compact, asymptotically flat manifold; otherwise $\chi$ would have a finite value and possible divergencies of the form $\chi/\varepsilon$ would have no reason to be ignored in the limit $\varepsilon\to 0$.
}
As it turns out, $\varepsilon = 0$ is a nonanalytic limit for the theory's Green function, implying that the limit $\varepsilon\to 0$ does not coincide with the theory defined in exactly two dimensions. Hence, for dimensional regularization to make sense one is forced to define two-dimensional gravity as the limit of $(2+\varepsilon)$-gravity and counterterms for the Lagrangian are therefore necessary even if it is topological.

Let us start by isolating the analytic part of the unrenormalized Newton's constant $G_0$ for finite $\varepsilon$
\begin{align}
G_0(\varepsilon)\mu^\varepsilon = \overline{G}(\mu) + \sum_n b_n(\overline{G}(\mu))\varepsilon^{-n}\,,
\end{align}
with renormalization group equation
\begin{align}
\mu\frac{{\rm d}}{{\rm d}\mu}G(\mu) = \beta(G(\mu), \varepsilon)\,,
\end{align}
where we dropped the bar for the dimensionless Newton's constant. At leading order, one finds the beta function
\begin{align}
\beta(G, \varepsilon) 
	= \varepsilon G + b_1(G) - G\frac{\partial b_1}{\partial G}\,.
\end{align}
Further expanding the beta function for small values of $G$ we can expect the coefficient $b_1$ to have the generic structure
\begin{align}
b_1(G) = b G^2 + o(G^3)\,,
\end{align}
leading to
\begin{align}
\label{eq:2dBetaGeneric}
\beta(G, \varepsilon) = \varepsilon G-b G^2\,.
\end{align}
We notice that, if the coefficient $b$ is positive, the $\beta$-function for $G$ has two fixed points: one corresponds to the Gau{\ss}ian theory for vanishing Newton's coupling, while the other occurs for the value $G^*= \varepsilon/b$. Inspecting the form of eq.~\eqref{eq:2dBetaGeneric} we notice that $G^*$ is a ultraviolet fixed point of the renormalization group flow which constitute a viable candidate for an asymptoptically safe theory of gravity at finite $\varepsilon$ close to two dimensions. The value of the fixed point is perturbative as long as one treats $\varepsilon$ as a small parameter, bypassing a lot of issues concerning non-perturbative computations. Moreover, as we send $\varepsilon$ to zero and approach $d=2$ dimensions, $G^*$ decreases until it collides with the Gau{\ss}ian fixed point and the theory becomes asymptotically free.

The key feature for such a scenario to hold is that the coefficient $b$ is positive. At one-loop, integrating the fluctuations of the metric with a linear splitting over a background (see \ref{app:bkgFieldMeth}
and in
\ref{app:covComputationsHK} for more details on the general procedures)
and quantum matter fluctuations, one finds the value \cite{Capper:1974ic, Capper:1974ed, Brown:1976wc, Tsao:1977tj, Weinberg:1980gg}
\begin{align}
b = \frac{2}{3}\left(19+6N_V-\frac{1}{2}N_F-N_S\right)\,.
\end{align}
where $N_V$ is the number of gauge fields coupled to the theory, $N_F$ the number of Majorana fermions and $N_S$ the number of scalar fields. One can see that the the fixed point $G^*$ exists for pure gravity, but also as long as the number of vector bosons is enough to compensate the presence of fermions and scalars.

Before buying into the asymptotic safety scenario, however, we need to pay special attention to the physical interpretation of our result. In a purely gravitational setting, Einstein equations set $R=0$ and Newton's coupling turns out to be inessential and there is no need fo it to reach a fixed point as $\mu$ grows to infinity. Upon the introduction of matter fields one can use Einstein equations to express $R$ in terms of the trace of the energy momentum tensor, therefore $G$ is not an independent coupling. However, the situation changes once we introduce new interactions to compare $\sqrt{g}R$ to. Two common choices are given by Gibbons-Hawking-York's boundary operator \cite{Gibbons:1976ue,York:1972sj,York:1986lje}, which is the most natural in the case of a manifold with boundary and was used in \cite{Gastmans:1977ad} to define the essential coupling of two dimensional gravity, and by the cosmological constant, which is useful in absence of boundary as shown in \cite{Christensen:1978sc}. One can fix either of these two operators using Einstein equations and extract the flow for an essential coupling expressed in terms of the curvature.

%%%%%%%%%%%%%%%%%%%%%%%%%%%%%%%%%%%%%%%%%%%%%%%%%%%%%%%%%%%%%
%%%%%%%%%%%%%%%%%%%%%%%%%%%%%%%%%%%%%%%%%%%%%%%%%%%%%%%%%%%%%
\section{The story of the conformal mode in $2+\varepsilon$}
\label{sec:JJandAK}
%%%%%%%%%%%%%%%%%%%%%%%%%%%%%%%%%%%%%%%%%%%%%%%%%%%%%%%%%%%%%
%%%%%%%%%%%%%%%%%%%%%%%%%%%%%%%%%%%%%%%%%%%%%%%%%%%%%%%%%%%%%

{
\it We discuss subtleties of the perturbative analysis
in $2+\varepsilon$ dimensions that involve the conformal mode
as they emerged in the original literature. These considerations were pivotal in pushing the analysis to two-loops,
because a nonlinear parametrization of the conformal mode of the metric was needed.
}

\smallskip

The application of dimensional regularization to two dimensional gravity was pursued with some dedication in a series of papers with the aim to go beyond the leading order discussed in Sect.~\ref{sect:2pluseps}.
The seminal attempts are from Kawai \& Ninomiya \cite{Kawai:1989yh}, and from Jack \& Jones \cite{Jack:1990ey}, who tried to renormalize the Einstein-Hilbert theory at two-loop in $d=2+\varepsilon$ dimensions with cosmological constant.\footnote{%
Notice that the original work by Jack \& Jones actually studies gravity in $d=2-\varepsilon$. It is worth pointing out that, formally, Feynman diagrams are indeed convergent \textit{below} $d=2$ rather than above. Moreover, the region $1<d<2$ has the advantage of be free of conformal instability. It is simple to formally continue the theory above $d=2$ \textit{after} regularizing Feynman integrals, though the continuation may encounter nonperturbative problems \cite{Martini:2021lcx}. Here we just present everything in $d=2+\varepsilon$ for a better comparison among all works in the literature.
}
The seminal papers \cite{Kawai:1989yh} and \cite{Jack:1990ey} draw very different conclusions on the renormalizability of gravity close to two dimensions.
We present the historical solution of the conundrum here,
while we reserve our modern take for Sect.~\ref{sec:2dRevised}.

The origin of the hurdle is in Ref.~\cite{Jack:1990ey},
where the authors point out possible issues with dimensional poles,
which, in the authors' interpretation,
hint that gravity is not two-loop renormalizable in $d=2+\varepsilon$.
In short, the problem is caused by the mixing of the dimensional poles coming from regulating Feynman diagrams with the kinematical poles coming from the fact that Einstein-Hilbert gravity is topological in $d=2$ (so the number of degrees of freedom changes discontinuously).
After that, in an impressive series of papers \cite{Kawai:1992np, Kawai:1993mb, Fukuma:1993np, Aida:1994zc, Nishimura:1994qh, Aida:1994np, Kawai:1995ju, Aida:1996zn} Kawai and many collaborators improved on the theoretical aspects of gravity close to two dimensions and its renormalization properties following the original results of \cite{Kawai:1989yh}.
Ultimately, these efforts led to the result of Aida \& Kitazawa \cite{Aida:1996zn} where the renormalizability of quantum gravity in $d=2+\varepsilon$ in the conformal gauge was proved to two-loop and the critical exponents of the theory were computed.

In this section we want to give a concise review of the difficulties encountered by Jack \& Jones and of the results obtained in the conformal gauge, mainly following \cite{Aida:1994zc, Aida:1996zn}. Some considerations will play an important role in the analysis of Sect.~\ref{sec:2dRevised}, which instead comes from more recent works \cite{Gielen:2018pvk, Martini:2021slj}. We start by reviewing the peculiar importance of the conformal mode and its relation to the fact that the number of degrees of freedom propagated by the Einstein-Hilbert action faces a discontinuity in $d=2$.

We start by following \cite{Jack:1990ey} and considering the action
\begin{align}
\label{eq:JJaction}
S[g] = -\int{\rm d}^dx\sqrt{g}\;\left(\frac{Z_G}{G}R-Z_{\Lambda}\Lambda\right)\,,
\end{align}
where $Z_G$ and $Z_\Lambda$ are renormalization constants. Consider now a splitting of the metric in a background and fluctuations as 
\begin{align}
g_{\mu\nu} = Z_g\left(\bar{g}_{\mu\nu} + Z_h\sqrt{G}h_{\mu\nu}\right)\,,
\end{align}
in which $Z_g$ and $Z_h$ are wavefunction renormalization for background and fluctuations respectively. Diffeomorphisms are gauge-fixed using the Feynman-de Donder gauge fixing action
\begin{align}
\label{eq:deDonderGaugeAction}
\begin{split}
S_{gf}[h;\bar{g}] &= \frac{1}{2}\int {\rm d}^d x \sqrt{\bar{g}}\;\bar{g}^{\mu\nu}F_\mu F_\nu\,,\\
F_\mu &= \bar{\nabla}_\alpha h^\alpha_\mu -\frac{1}{2}\bar{\nabla}_\mu h^\alpha_\alpha\,.
\end{split}
\end{align}
The part of the gauge fixed action that is quadratic in the fluctuations and covariant over the background, i.e., the Hessian is
\begin{align}
\label{eq:EinsteinHessian}
S^{(2)}[h;\bar{g}] = \frac{1}{2}\int {\rm d}^d x\sqrt{\bar{g}}\;h_{\mu\nu}
 \left(-\bar{\nabla}^2 K^{\mu\nu\alpha\beta}+X^{\mu\nu\alpha\beta}\right)
 h_{\alpha\beta}\,,
\end{align}
where $\bar{\nabla}^2 = \bar{\nabla}^{\mu}\bar{\nabla}_\mu$ is the background compatible covariant Laplacian.
The endomorphism $X$ is a function of background curvature operators and the matrix $K$ is
\begin{align}
K^{\mu\nu\alpha\beta} = \frac{1}{2}\left(\bar{g}^{\mu\alpha}\bar{g}^{\nu\beta}+\bar{g}^{\mu\beta}\bar{g}^{\nu\alpha}-\bar{g}^{\mu\nu}\bar{g}^{\alpha\beta}\right)\,.
\end{align}
The differential operator appearing in \eqref{eq:EinsteinHessian} is not invertible in $d=2$ because the matrix $K$ is not invertible.
In general $d$ this problem takes the form of a divergent term for $d\sim 2$
in the inverse
\begin{align}
K^{-1}_{\mu\nu\alpha\beta} = \frac{1}{2}\left(\bar{g}_{\mu\alpha}\bar{g}_{\nu\beta}+\bar{g}_{\mu\beta}\bar{g}_{\nu\alpha}-\frac{1}{d-2}\bar{g}_{\mu\nu}\bar{g}_{\alpha\beta}\right)\,.
\end{align}
The pole in $K^{-1}$ is of kinematical origin because it is caused by the fact that in (and only in) $d=2$ the only degree of freedom of the metric that can propagate is the conformal one.

However, before taking the limit $\varepsilon\to 0$, we could still apply the heat kernel techniques to compute the effective action to the Hessian \eqref{eq:EinsteinHessian} multiplied by $K^{-1}$, i.e., consider the kinetic operator
\begin{align}
-\nabla^2 I_{\mu\nu\alpha\beta}+K^{-1}_{\mu\nu\rho\sigma}{X^{\rho\sigma}}_{\alpha\beta}\,,
\end{align}
that is finite for $d\sim2$, given $I$ the identity on the space of symmetric tensors. This operation can be thought of as a change of variable with unit Jacobian in the path integral, which should not affect the one-loop computation of the effective action, however divergences of higher loop diagrams are potentially sensitive to the redefinition caused by $K^{-1}$. In fact, it seems that all the propagators of loop diagrams will come with an extra (kinematical) $\varepsilon$-pole that must combine with similar poles coming from dimensional regularization. The authors of Ref.~\cite{Jack:1990ey} then deduce that subleading divergences will fail to cancel (the cancellation of subleading divergences beyond one-loop is always a delicate point, specifically they show that one cannot find a wave function renormalization that cancels the nonlocal divergences at two-loop).

However, the presence of the kinematical pole does not find its origin in the dimensional regularization method, rather in the symmetries of the original theory and in the topological nature of the scalar curvature in $d=2$. One can see that the role of the $(d-2)^{-1}$ term is to give an infinite weight to the trace mode of the fluctuations as $d$ approaches $2$ and to flip the sign of the kinetic term for such mode.

To better understand the situation, let us consider the Einstein-Hilbert action without cosmological constant and parametrize the full metric, which we now refer to as $\bar{g}$, in terms of a fiducial metric $\hat{g}$ and the conformal mode $\phi$ as
\begin{align}
g_{\mu\nu}= \hat{g}_{\mu\nu}e^{-\phi}\,.
\end{align}
One gets
\begin{align}
\int{\rm d}^d x \sqrt{g}\;R =
 \int {\rm d}^dx\sqrt{\hat{g}}\;e^{-\frac{\varepsilon}{2}\phi}\left[\hat{R}-\varepsilon\frac{d-1}{4}\hat{g}^{\mu\nu}\partial_\mu\phi\partial_\nu\phi\right]\,,
\end{align}
from which is manifest that the action is Weyl invariant in the limit $\varepsilon\to 0$ as expected by its the topological nature. However, if we now parametrize the conformal mode as
\begin{align}
\label{eq:dilatonRescaled}
e^{-\frac{\varepsilon}{4}\phi} = \sqrt{\frac{\varepsilon}{8(d-1)}}\psi\,,
\end{align}
it is easy to see that the original action is equivalent to the theory of a scalar conformally coupled to the background geometry
\begin{align}
\label{eq:actionAKOriginal}
\int{\rm d}^d x \sqrt{g}\;R = \int {\rm d}^dx \sqrt{\hat{g}}\;
	\left[\hat{R}\frac{\varepsilon}{8(d-1)}\psi^2-\frac{1}{2}\hat{g}^{\mu\nu}\partial_{\mu}\psi\partial_{\nu}\psi\right]\,.
\end{align}
We can then try to quantize the latter action under the prescription of preserving Weyl invariance.\footnote{%
This procedure involves a rescaling in the dilaton mode that is singular
as seen in \eqref{eq:dilatonRescaled}. The interpretation reviewed in Sect.~\ref{sec:2dRevised} solves the problem in an entirely different way.
}
Note that the domain of the field $\psi$ is classically constrained to be positive, according to Eq.~\eqref{eq:dilatonRescaled}, but in the quantization procedure all fluctuations are allowed.

An important point to keep in mind is that we are enlarging the symmetry group of our theory by including Weyl symmetry. However, it so happens that, locally, the diffeomorphisms group ${\rm Diff}(\mathcal{M})$ of the spacetime manifold $\mathcal{M}$ is isomorphic to the semidirect product ${\rm Diff}^*={\rm SDiff}(\mathcal{M})\ltimes{\rm Weyl}$, where ${\rm SDiff}$ is the group of volume preserving diffeomorphisms, while ${\rm Weyl}$ is the group of Weyl transformation of the metric.
In terms of the classical fields $\hat{g}$ and $\phi$, ${\rm Diff}^*$ acts as
\begin{align}
\delta^*_\xi \hat{g}_{\mu\nu} = \mathcal{L}_\xi \hat{g}_{\mu\nu} 
 - \frac{2}{d}\hat{g}_{\mu\nu}\hat{\nabla}_\rho\xi^\rho\,,
 \qquad
\delta^*_\xi\phi =&\mathcal{L}_\xi\phi + \frac{\varepsilon}{2d}\phi\hat{\nabla}_\rho\xi^\rho\,,
\end{align}
where $\mathcal{L}_\xi$ is the Lie derivative along the vector field $\xi$ and $\hat{\nabla}$ is the covariant derivative compatible with $\hat{g}$.
In the background field method one can break ${\rm Diff}(\mathcal{M})\ltimes{\rm Weyl}$ to ${\rm Diff}^*$ by splitting the fiducial metric as
\begin{align}
\label{eq:expPar}
\hat{g}_{\mu\nu} = \bar{g}_{\mu\rho}\left(e^h\right)^\rho_\nu
\end{align}
and requiring that the fluctuation field is traceless ${g}^{\mu\nu}h_{\mu\nu} = 0$ so that $\det\hat{g}=\det g$. One then expands the scalar field around a constant vev as
\begin{align}
\sqrt{\frac{\varepsilon}{8(d-1)}}\psi\rightarrow \mu^{\frac{\varepsilon}{4}}
 \left(1+\frac{1}{2}\sqrt{\frac{\varepsilon}{2(d-1)}}\psi\right)\,,
\end{align}
and considers the gauge fixing action
\begin{align}
\label{eq:gaugeFixAK1}
\begin{split}
& S_{gf}[h, \psi; \bar{g}] = \frac{1}{2}\int{\rm d}^dx\sqrt{\bar{g}}\;
		\bar{g}^{\mu\nu}F_\mu F_\nu\,,\\
 & F_\mu = \bar{\nabla}^\alpha h_{\alpha\mu}-\sqrt{\frac{\varepsilon}{2(d-1)}}\partial_{\mu}\psi\,.
 \end{split}
\end{align}
Upon including the ghost action in the usual way, we can compute the one-loop divergence for the effective action of the theory and obtain
\begin{align}
\label{eq:weylAnomaly}
\Gamma_{\infty} = \frac{25}{24\pi}\frac{\mu^\varepsilon}{\varepsilon}\int{\rm d}^d x
 \sqrt{\bar{g}}\;\bar{R}\,.
\end{align}
Eq.~\eqref{eq:weylAnomaly} breaks Weyl invariance and makes the quantum theory anomalous.

At this point one can resort on two strategies to obtain a Weyl symmetric quantum theory. The first option is to add matter degrees of freedom hoping that the corresponding counterterms can be used to cancel the anomaly. In this approach the most popular choice is to include conformally coupled scalar fields $\varphi_i$ at tree level as
\begin{align}
\begin{split}
S =&\frac{\mu^\varepsilon}{G} \int{\rm d}^d x\sqrt{\bar{g}}\;\left\{\hat{R}\left[\left(1+\frac{1}{2}
 \sqrt{\frac{\varepsilon}{2(d-1)}}\psi\right)^2
 -\frac{\varepsilon}{8(d-1)}\varphi_i^2\right]\right.\\
&\left.-\frac{1}{2}\partial_\mu\psi\partial_\nu\psi \hat{g}^{\mu\nu}
 +\frac{1}{2}\partial_\mu\varphi_i\partial_\nu\varphi_i \hat{g}^{\mu\nu}\right\}\,.
 \end{split}
\end{align}
Equation \eqref{eq:weylAnomaly} turns into
\begin{align}
\label{eq:AK1loop}
\Gamma_{\infty} = \frac{25-c}{24\pi}\frac{\mu^\varepsilon}{\varepsilon}\int{\rm d}^d x 
 \sqrt{\bar{g}}\;\bar{R}\,,
\end{align}
where $c$ is the number of scalar fields. One can then make the theory anomaly free by choosing $c=25$. This way of restoring Weyl symmetry has been popularized by string theory, where the scalar multiplet $\{\psi,\,\varphi_i\}$ is interpreted as the embedding map of the string into the target space. In the string's language, the coefficient $25-c$ is actually rewritten as $26-(c+1)$, where $1$ is for the string's conformal mode, which does not propagate over the Polyakov's string, and $26$ is critical the dimension of the string's target space.

Alternatively, the cancellation of the anomaly can be achieved if we include in the bare action a term of the form
\begin{align}
S_t = q \int{\rm d}^d x\sqrt{\bar{g}}\; \hat{R}\psi\,.
\end{align}
The constant $q$ is known as the ``topological'' charge and does not renormalize in (and only in) $d=2$ \cite{Martini:2021lcx}. Hence, even though $q$ breaks Weyl symmetry at tree level, we are free to choose it so that it cancels the quantum anomaly. We return on these aspects of the conformal gauge of two dimensional gravity in Sect.~\ref{sec:2dRevised}.

%%%%%%%%%%%%%%%%%%%%%%%%%%%%%%%%%%%%%%%%%%%%%%%%%%%%%%%%%%%%%
%%%%%%%%%%%%%%%%%%%%%%%%%%%%%%%%%%%%%%%%%%%%%%%%%%%%%%%%%%%%%
\subsection{Two-dimensional gravity at two-loops}
%%%%%%%%%%%%%%%%%%%%%%%%%%%%%%%%%%%%%%%%%%%%%%%%%%%%%%%%%%%%%
%%%%%%%%%%%%%%%%%%%%%%%%%%%%%%%%%%%%%%%%%%%%%%%%%%%%%%%%%%%%%

The leading loop computation is not enough to guarantee the renormalizability of the theory because the cancellation of higher loops is delicate for the case of gravity.
The two-loop computation for gravity in $d=2+\varepsilon$ dimensions using the rescaling of the conformal mode was performed by Aida \& Kitazawa \cite{Aida:1996zn}. Given the uniqueness and the technical complexity of the two-loop computation, we recommend that it is repeated in the future by some corageous colleague.

For completeness, Aida \& Kitazawa consider a system of gravity coupled to matter fields and study the renormalization of the gravitational sector
\begin{align}
S_{\text{grav}} =& \frac{\mu^{\varepsilon}}{G}\int {\rm d}^d x \sqrt{\bar{g}}\;
  \left\{\hat{R}\left(1 + a\psi + \varepsilon b \psi^2 \right) 
  -\frac{1}{2}\hat{g}^{\mu\nu}\partial_\mu\psi\partial_\nu\psi\right\}\,,\\
S_{\text{matter}} =& \frac{\mu^{\varepsilon}}{G}\int {\rm d}^d x \sqrt{\bar{g}}\;
  \left\{\frac{1}{2}\hat{g}^{\mu\nu}\partial_\mu\varphi_i\partial_\nu\varphi_i
  -\varepsilon b \varphi_i^2 \hat{R}\right\}\,,
\end{align}
where we adopted the same conventions as the previous section for the parametrization of metric degrees of freedom. Classically, the conformal values of the couplings $a$ and $b$ are given by $a^2 = 4\varepsilon b = \varepsilon/2(d-1)$ and are understood as the bare values of the microscopic (bare) theory. At one-loop, $a$ receives radiative corrections as long as $d\neq 2$ and has a Gau{\ss}ian fixed point. Here, the gauge symmetry of the fields reads
\begin{align}
\begin{split}
\delta^*_\xi \hat{g}_{\mu\nu} =& \mathcal{L}_\xi \hat{g}_{\mu\nu} 
		- \frac{2}{d}\hat{g}_{\mu\nu}\hat{\nabla}_\rho\xi^\rho\,,\\
\delta^*_\xi \psi =& \mathcal{L}_\xi\psi 
		+ \frac{2}{d}\left[(d-1)a+\frac{\varepsilon}{4}\psi\right]\hat{\nabla}_\rho\xi^\rho\,,\\
\delta^*_\xi \varphi_i =& \mathcal{L}_\xi\varphi_i
		+ \frac{2}{d}\left(\frac{\varepsilon}{4}\varphi_i\right)\hat{\nabla}_\rho\xi^\rho\,.
\end{split}
\end{align}
The gauge fixing action in this parametrization can be obtained as a slight modification of \eqref{eq:gaugeFixAK1}
\begin{align}
S_{gf} = \frac{\mu^\varepsilon}{G}\int{\rm d}^d x\sqrt{\bar{g}}\; \frac{1}{2}
		\left(\bar{\nabla}_\mu h^\mu_\nu-a\partial_\nu\psi\right)\left(\bar{\nabla}_\rho h^{\rho\nu}-a\partial^\nu\psi\right)\,.
\end{align}

Evaluating the two-loop diagrams of the theory one can explicitly observe the cancellation of $\frac{1}{\varepsilon^2}$ poles. Most of the nonlocal divergences arising at two-loop level are canceled by diagrams containing the one-loop counterterms, the only surviving one being proportional to the equations of motion. The total divergence of the theory is given by \cite{Aida:1996zn}
\begin{align}
\label{eq:aidaKitazawa2loopDiv}
\begin{split}
\Gamma_\infty =& \frac{G}{(4\pi)^2}\int{\rm d}^d x\sqrt{\bar{g}}\;\frac{-79+3c}{8\varepsilon}\bar{R}\\
& + \frac{G}{4\pi}\int{\rm d}^dx\sqrt{\bar{g}}\;\left(-\frac{11}{12\varepsilon}\right)
		\left(\bar{R}_{\mu\nu}-\frac{1}{2}\bar{R} \bar{g}_{\mu\nu}\right){{\overline{G}^\mu}_\rho}^{\nu\rho}\,,
\end{split}
\end{align}
where $\overline{G}^{\mu\nu\rho\sigma}$ is defined through the Seeley-de Witt expansion of the graviton propagator (see Eqs.~\eqref{eq:green-sdw-expansion} and \eqref{eq:green-sdw-expansion-rewritten} in \ref{app:covComputationsHK}) as
\begin{align}
G^{\mu\nu\rho\sigma} (x, x') = G_0(x, x')a_0^{\mu\nu\rho\sigma}(x, x') + 
		G_1(x, x')a_1^{\mu\nu\rho\sigma}(x, x') + \overline{G}^{\mu\nu\rho\sigma}\,.
\end{align}
$\overline{G}$ is regular in the coincident limit but depends on global aspects of the manifold and its cancellation from physical divergences is crucial for the renormalization program. In our case $\overline{G}$ only appears multiplying Einstein's tensor meaning that it can be eliminated by going on-shell on the background geometry (in the limit $\varepsilon\to 0$).
Alternatively, as shown in \cite{Aida:1994zc, Aida:1996zn}, one can remove the nonlocal two-loop divergence with a suitable wave function renormalization $Z_h$ with $O(\varepsilon^{-1})$ poles. In this case, the parametrization of $Z_h$ requires nonlinear terms in the fluctuations $h_{\mu\nu}$.
The final two-loop divergence of gravity without cosmological constant in the conformal gauge is
\begin{align}
\Gamma_\infty =& \frac{G}{(4\pi)^2}\int{\rm d}^d x\sqrt{\bar{g}}\;\frac{3c-79}{8\varepsilon}\bar{R}\,.
\end{align}

%%%%%%%%%%%%%%%%%%%%%%%%%%%%%%%%%%%%%%%%%%%%%%%%%%%%%%%%%%%%%
%%%%%%%%%%%%%%%%%%%%%%%%%%%%%%%%%%%%%%%%%%%%%%%%%%%%%%%%%%%%%
\subsection{The renormalization of the cosmological constant}
%%%%%%%%%%%%%%%%%%%%%%%%%%%%%%%%%%%%%%%%%%%%%%%%%%%%%%%%%%%%%
%%%%%%%%%%%%%%%%%%%%%%%%%%%%%%%%%%%%%%%%%%%%%%%%%%%%%%%%%%%%%

One can include the cosmological constant as a perturbation to the theory and renormalize it as a composite operator. We have
\begin{align}
\begin{split}
\Lambda \int{\rm d}^d x\sqrt{g} =& \Lambda \int{\rm d}^d x\sqrt{\bar{g}}\;
		\left(1+\frac{2\varepsilon b}{a}\psi\right)^{\frac{2d}{\varepsilon}}\\
\simeq& \Lambda \int{\rm d}^dx\sqrt{\bar{g}}\;
		\exp\left\{\left(1-\frac{\varepsilon}{2}\right)\frac{\psi}{a}
		-\frac{\varepsilon}{8a^2}\psi^2+\dots\right\}\,.
\end{split}
\end{align}

Keeping the quadratic order in $\psi$ and computing the diagrams with one insertion of the composite the only diagrams that are relevant are those with $\psi$ propagators. One finds for the divergent parts of the effective action the following extra poles at $1$- and $2$-loop
\begin{align}
\begin{split}
\Gamma_\Lambda^1 =& -\int{\rm d}^dx\sqrt{\bar{g}}\;
		\frac{G\Lambda}{4\pi}\left(\frac{1}{a\varepsilon}+\frac{1}{\varepsilon}\right)\;\\
\Gamma_\Lambda^2 =& \int{\rm d}^dx\sqrt{\bar{g}}\; 
		\left[-\frac{G^2\Lambda}{16\pi^2}\left(\frac{1}{4a^2\varepsilon}
		+\frac{3}{\varepsilon^2}+\frac{35}{8\varepsilon}\right)
		+\frac{G^2\Lambda}{4\pi\varepsilon}{\overline{G}_{\mu\nu}}^{\mu\nu}
		+\frac{G}{4\pi\varepsilon}a^2\bar{\nabla}^2\overline{G}_{\psi\psi}\right]\,,
\end{split}
\end{align}
where $\overline{G}_{\psi\psi}$ contains nonlocalities due to the propagator of $\psi$. As we can see from the structure of the two-loop divergence, we have two nonlocal divergencies and an $\varepsilon^2$ pole. Once more, the term proportional to ${\overline{G}_{\mu\nu}}^{\mu\nu}$ can be removed by employing a particular wave function renormalization for $\psi$ with nonlinear terms in $h_{\mu\nu}$.
The term proportional to $\overline{G}_{\psi\psi}$ is due to one-loop subdivergences of the ghost and graviton fields, contributing to the divergent part of the kinetic term for $\psi$. We can get rid of this by including the  counterterm in the one-loop action
\begin{align}
\label{eq:AK1loopExtra}
\frac{(d-1)a^2}{2\pi\varepsilon}\int{\rm d}^dx\sqrt{\bar{g}}\;
		\left[\hat{R}\left(1+a\psi+\varepsilon b\psi^2\right)
		-\frac{1}{2}\partial_\mu\psi\partial^\mu\psi\right]\,.
\end{align}
Finally, the $\varepsilon^2$ pole is removed exploiting one's freedom to rescale the background metric by a factor
\begin{align}
Z_{bg} = 1 - \left(\frac{G}{4\pi\varepsilon}\right)^2\,.
\end{align}
As a consequence, we get a rescaling of the Einstein-Hilbert action according to
\begin{align}
S_{EH}\longrightarrow Z_{bg}^\frac{\varepsilon}{2}\int{\rm d}^dx\sqrt{g}\;R\,.
\end{align}
All these final manipulations do not add extra divergences to the purely gravitational part discussed in the previous section. The final divergence for two-loop quantum gravity in $d=2+\varepsilon$ dimension in the conformal gauge turns out to be
\begin{align}
\label{eq:2loopAKFinal}
\Gamma_\infty = -3\frac{G}{(4\pi)^2}\int{\rm d}^dx \sqrt{\bar{g}}\;\frac{25-c}{8\varepsilon}\bar{R}\,.
\end{align}
We notice that eq.~\eqref{eq:2loopAKFinal} vanishes for $c=25$, which corresponds to the critical string.
From \eqref{eq:AK1loop}, \eqref{eq:AK1loopExtra} and \eqref{eq:2loopAKFinal} one can extract the beta function for the Newton's coupling
\begin{align}
\beta_G = \varepsilon G - \frac{25-c}{24\pi}G^2-5\frac{25-c}{48\pi^2}G^3\,.
\end{align}
The simple poles appearing in the renormalization of the cosmological constant and proportional to $\frac{1}{a^2}$ are due to the singular vev for $\psi^2$
\begin{align}
\langle\psi^2\rangle = \frac{G}{2\pi\varepsilon}\left(1+\frac{G}{16\pi}\right)\,.
\end{align}
These poles can become important close to the UV fixed point, where $a$ vanishes with the renormalization flow. One can perform a resummation of such contributions to obtain the following anomalous dimension for the volume operator
\begin{align}
\gamma_\Lambda = 2-\frac{G}{8\pi}+\frac{8\pi a^2}{G}-\frac{8\pi}{G}\sqrt{a^4+\frac{Ga^2}{2\pi}
		+\frac{1}{2}\left(\frac{Ga}{4\pi}\right)^2}\,.
\end{align}
In the deep ultraviolet, one has the fixed point $G^* = \frac{24\pi\varepsilon}{25-c}+o(\varepsilon^2)$ and $a$ vanishes, therefore we may wright
\begin{align}
\gamma^*_\Lambda = 2\left(1-\frac{G^*}{16\pi}\right)\,.
\end{align}
However, the physical content lies in the scaling relation between $G$ and $\Lambda$ which, in the UV, is given by
\begin{align}
\left(\frac{1}{G}-\frac{1}{G^*}\right)^{\varepsilon+\frac{3\varepsilon}{25-c}}
		\sim\Lambda^\varepsilon\,,
\end{align}
as opposed to the scaling at the IR Gau{\ss}ian fixed point that reads
\begin{align}
\frac{1}{G^d}\sim\Lambda^\varepsilon\,.
\end{align}

%%%%%%%%%%%%%%%%%%%%%%%%%%%%%%%%%%%%%%%%%%%%%%%%%%%%%%%%%%%%%
%%%%%%%%%%%%%%%%%%%%%%%%%%%%%%%%%%%%%%%%%%%%%%%%%%%%%%%%%%%%%
\section{Revisiting quantum gravity in $2+\varepsilon$}
\label{sec:2dRevised}
%%%%%%%%%%%%%%%%%%%%%%%%%%%%%%%%%%%%%%%%%%%%%%%%%%%%%%%%%%%%%
%%%%%%%%%%%%%%%%%%%%%%%%%%%%%%%%%%%%%%%%%%%%%%%%%%%%%%%%%%%%%

{\it We discuss the insights coming from a modified renormalization scheme with different metric parameterizations. The main idea is to treat separately the dimensionality of spacetime that counts the degrees of freedom of the metric
from the dimensionality that is analytically continued to regulate the covariant Feynman diagrams.}

\smallskip

The success of the renormalization program for two dimensional quantum gravity in the conformal gauge rises several questions. 
First of all, one would expect the two formulations of quantum gravity to contain the same physics on the ground that the two forms of the microscopic action can be seen as two different but equivalent classical frames. However the one-loop beta functions seem to have very different phenomenology due to a different central charge value.
One may be tempted to attribute such difference to different universality classes since one-loop beta functions are expected to be scheme independent, however it is unclear why a simple frame transformation would cause such a dramatic change in the path integral. A possible speculation is that the parametrization of the fluctuations could be the cause of the different results, because the linear difference $h_{\mu\nu} = g_{\mu\nu}-\overline{g}_{\mu\nu}$ parametrizes an integration domain that differs substantially in the space of metrics (unless imposing nonlinear Ward identities for the splitting symmetry, see \ref{app:bkgFieldMeth}), as opposed to the exponential parametrization that is used in the conformal gauge, which preserves the integration domain by construction (i.e., constraining the metric to have a fixed signature).
Moreover, the difference in higher loop computations suggests that a more careful analysis of the degrees of freedom must apply, keeping in mind also the importance of the path integral measure.

Keeping in mind the original idea from Weinberg, recent works \cite{Falls:2015qga, Nink:2015lmq, Falls:2017cze, Martini:2021slj, Martini:2021lcx} revived the computations of gravity in $d=2+\varepsilon$ dimensions trying to pay special attention to the scheme dependence of the renormalization flow and the parametrization of the symmetry group. As mentioned in section \ref{sec:JJandAK}, the kinematic pole as $d\to 2$ in the tree-level propagator does not originate due to our regularization scheme. It seems natural, also in light of the conformal gauge results, to keep the kinematic pole $(d-2)^{-1}$ explicit, without \emph{a priori} identifying it with the (inverse of the) $\varepsilon$ variable of dimensional regularization. The basic idea behind such a choice is that $d$, whenever entering a tensorial structure, parametrizes the gauge symmetry of gravity, much like the parameter $N$ does in $SU(N)$ Yang-Mills theories. However, at the same time, when one is computing Feynman diagrams, the integration measure will be fictitiously shifted away from its critical dimension $d_c=2$ with the only purpose of making the diagrams convergent, and this causes divergence of the theory to isolate in the form of $\varepsilon$-poles. The two ``instances'' of the dimension of spacetime can be equated later, after the theory has been properly renormalized, in order to extract results in some specific dimension. Finally, if one hopes to find a true physical meaning in the renormalized coupling constants of the theory, it is of crucial importance to identify the gauge and parametrization independent counterterms.

Computing the one-loop effective action using the regularization scheme described above from the action \eqref{eq:JJaction}, one can inspect the dependence on the background splitting of the metric parametrizing it as
\begin{align}
\label{eq:linearSplittingWithLambda}
g_{\mu\nu} = \bar{g}_{\mu\nu} + h_{\mu\nu}
		+\frac{\lambda}{2}h_{\mu\rho}\bar{g}^{\rho\theta}h_{\theta\nu}\,,
\end{align}
where we truncate the splitting to the quadratic order for simplicity,
but further powers can in principle be included.
Note that the symbol $\bar{g}$ now represents a different background with respect to the previous section, because here we do not isolate the conformal mode.
The dependence on the gauge choice can be made explicit by including a deformation of the Feynman-de Donder gauge fixing \eqref{eq:deDonderGaugeAction}
\begin{align}
\begin{split}
S_{gf}[h;\bar{g}] &= \frac{1}{2}\int {\rm d}^d x \sqrt{\bar{g}}\;\bar{g}^{\mu\nu}F_\mu F_\nu\,,\\
F_\mu =& \bar{\nabla}_\alpha h^\alpha_\mu -\frac{1+\delta\beta}{2}\bar{\nabla}_\mu h^\alpha_\alpha\,.
\end{split}
\end{align}
The deformation parameter $\delta\beta$ is chosen such that $\left|\delta\beta\right|\ll1$, also for simplicity.\footnote{%
To elaborate on a subtle point: in a perturbative approach one could think of $\delta\beta$ as a new coupling and ``renormalize'' the gauge fixing deformation as a composite operator, because it enters as an operator insertion in the loops when considering the leading order in $\delta\beta$.
In practice, the independence of the final physical results from $\delta\beta$
tests that their derivative with respect to the gauge parameter is zero. It does not test that they are fully gauge-independent, although it provides a strong indication.
}
In the following we consider only the leading order correction due to $\delta\beta$.
We get\footnote{%
As before, Feynmann integrals are formally computed below $d=2$, but we report the result continued above two dimension.
}
\begin{align}
\label{eq:martZanuEinstDiv}
\Gamma_{\infty} = -\frac{\mu^{\varepsilon}}{\varepsilon}\int {\rm d}^d x \sqrt{\bar{g}}\;\left[
		A \bar{R} +J_{\mu\nu}\left(G^{\mu\nu}+\frac{1}{2}\lambda G \bar{g}^{\mu\nu}\right)\right]\,,
\end{align}
where $G^{\mu\nu}=\bar{R}^{\mu\nu}-\frac{1}{2}\bar{R}\bar{g}^{\mu\nu}$ is the Einstein's tensor for the background metric \cite{Martini:2021slj}. As for the coefficients, we have
\begin{align}
\begin{split}
A =& \frac{36+3d-d^2}{48\pi}\,,\\
J_{\mu\nu} =& \frac{\bar{g}_{\mu\nu}}{4\pi}\left[\frac{d^2-d-4}{2(d-2)}\lambda
		-\delta\beta\left(2+\frac{2\lambda}{d-2}\right)-d-1\right]\,,
\end{split}
\end{align}
We notice, as expected, that going on-shell eliminates both the parametrization and gauge dependences, because $J_{\mu\nu}$ decouples. Moreover, the possible extra poles of kinematical origin appear only in combination with the parameter $\lambda$ and also disappear on-shell.

The role of $J_{\mu\nu}$ is that of a wave function renormalization that needs to be included in higher loop computations. At subleading order in the renormalization with this scheme, which are not available yet, the source $J_{\mu\nu}$ must be used to dress external graviton lines,
and thus produces a notion of ``quantum metric'', which is the natural argument of the effective action that is also fully dressed by quantum corrections.
Strictly speaking, since the metric is not an observable, it is not strange that its renormalization carries the parametric and gauge dependencies. Although this does not prove renormalizability of Einstein-Hilbert formulation in $d=2+\varepsilon$, it suggests that higher loop computation might turn out to be less pathological than those expected in Ref.~\cite{Jack:1990ey}.

It is important to notice how the beta function for the Newton's constant
\begin{align}
\label{eq:betaMartZanuEinst}
\beta_G = \varepsilon G - \frac{36+3d-d^2}{48\pi}G^2\,,
\end{align}
is parametrically finite in any $d$ and gauge-independent, with a UV interacting fixed point of order $\varepsilon$ that disappears in dimension $d_c \approx 7.7$ \cite{Falls:2015qga, Martini:2021slj}, which contains the physical case of four dimensions. However, the result is perturbative and it is hard to expect that its predictions could be quantitatively accurate. Despite the limitations, this is one important result in that it finds that the gravitational fixed point exists within a \emph{conformal window}, which is different from the naive prediction the functional renormalization group in the background field \cite{Litim:2003vp}. In the limit $d\to 2$ the beta function \eqref{eq:betaMartZanuEinst} reproduces the ``central charge'' value, $c=-19$
of Ref.~\cite{Jack:1990ey}.

One can use the same prescription to investigate the formulation of gravity invariant under ${\rm Diff}^*$. As we pointed out in Sect.~\ref{sec:JJandAK}, the best way to parametrize fluctuations of the metric is through formula \eqref{eq:expPar}, so that the background becomes unimodular.
In this case one has multiple choices to impose the on-shell condition. One natural choice is to use the background equations of motion for the dilaton field, denoted $e$ in the following, and parametrize the divergent part of the theory.
Starting from the background action (without cosmological constant)
\begin{align}
\label{eq:actionMZAK}
S[\bar{g}, \psi] = -\frac{1}{G}\int {\rm d}^d x \sqrt{\bar{g}}\;\left[\psi^2 \bar{R} +4\frac{d-1}{\varepsilon}\bar{g}^{\mu\nu}\partial_\mu\psi\partial_\nu\psi\right]
\end{align}
with $\bar{g}$ a unimodular metric, the equations of motion read
\begin{align}
e \equiv 8\frac{d-1}{\varepsilon G}\bar{\nabla}^2\psi-\frac{2}{G}\bar{R}\psi = 0\,.
\end{align}
Again, we include a deformation of the gauge fixing \eqref{eq:gaugeFixAK1} that now reads\footnote{Note the different normalization of $\psi$ already evident comparing \eqref{eq:actionMZAK} and \eqref{eq:actionAKOriginal}.}
\begin{align}
\begin{split}
& S_{gf}[h, \psi; \bar{g}] = \frac{1}{2}\int{\rm d}^dx\sqrt{\bar{g}}\;
		\bar{g}^{\mu\nu}F_\mu F_\nu\,,\\
 & F_\mu =\psi_0 \bar{\nabla}^\alpha h_{\alpha\mu}-2(1+\delta\beta)\partial_{\mu}\psi\,,
\end{split}
\end{align}
where $\psi_0$ is the (non-constant) background for the conformal mode.
 
The divergent part of the effective action in the background field method is then given by
\begin{align}
\Gamma_{\infty} = -\frac{\mu^\varepsilon}{\varepsilon}\int {\rm d}^dx\sqrt{\bar{g}}\;
		\left[B\psi_0^2 \bar{R} + J\psi_0 e\right]\,,
\end{align}
where $\bar{R}$ is the Ricci scalar for the unimodular background. The coefficients $B$ and $J$ are given by
\begin{align}
\begin{split}
B =& -\frac{11d^4-44d^3-78d^2+180d-72}{96\pi d(d-1)}\\
&-\frac{(d-2)(18d^5-35d^4-132d^3+152d^2+48d-48)}{192\pi d^3(d-1)}\delta\beta\,,\\
J =& -\frac{3d^3-6d^2-12d+16}{8\pi d}-\frac{(d-2)(3d^2-4)(2d^2+d-2)}{16\pi d^3}\delta\beta\,.
\end{split}
\end{align}
The first thing to notice is that a residual gauge dependence for the on-shell counterterm survives if $d \neq 2$. This is not surprising since the Weyl symmetry used to construct the gauge group is not even realized at tree level for arbitrary dimension. In this case, in fact, it is difficult to talk about ``Weyl anomaly'' as the Weyl symmetry was not there to begin with. In the limit $d\to 2$ the residual gauge dependence vanishes on-shell and one recovers the beta function
\begin{align}
\beta_G = -\frac{25}{24\pi}G^2\,.
\end{align}
However, while \eqref{eq:betaMartZanuEinst} was obtained by fixing $\Lambda$ and measuring the flow of $G$ with respect to the volume operator, here we have not yet considered the cosmological constant. In fact, the use of the conformal mode to impose the on-shell condition boils down to a different scheme and a different choice of essential couplings. To have a proper comparison, one can study the renormalization of the volume operator as a composite of the dilaton field given by
\begin{align}
V[\psi] = \Lambda \int{\rm d}^dx\sqrt{\bar{g}}\;\psi^{\frac{2d}{\varepsilon}}\,,
\end{align}
and impose the relation
\begin{align}
\Lambda = \frac{\psi^{-\frac{d+2}{d-2}}}{d G}\left[(d-2)\bar{R}\psi -4(d-1)\bar{\nabla}^2\psi\right]\,.
\end{align}
In so doing one recovers, in the limit $d\to 2$, the beta function \eqref{eq:betaMartZanuEinst} for the Newton's constant. This reveals that the two results for the values of the central charge are actually due to a difference in the scheme used to go on-shell. Consequently, the value $c=25$ comes from a parametrization of the group of diffeomorphisms that contains the Weyl group \textit{and} from preserving the subgroup generated by Weyl transformations in the use of the equations of motions, but this is possible \emph{only} in $d=2$. 

As a final remark, it is interesting to notice that, in $d=2$, one way to realize a Weyl-symmetric quantum theory is to first break explicitly the symmetry at tree level with a term of the form
\begin{align}
\Delta S_t[g, \psi] = q \int {\rm d}^2x \sqrt{g}\;R\psi\,.
\end{align}
The running of the coupling $q$, known occasionally as topological charge, under the renormalization flow is trivial (only in $d=2$), therefore one can set it to any value that cancels the one-loop Weyl anomaly induced by the other terms. However, using the prescription explained above one can explicitly check that $q$ actually renormalizes for $d>2$ and so it is not useful to restore the invariance. Again, this is compatible with the fact that the symmetry is well defined only in $d=2$.

All these computations show that, although unimodular dilaton gravity is well-suited for applications to two dimensional gravity and string theory (where the anomaly can be cancelled through the topological charge), most of its nice properties do not hold beyond $d=2$ and is not a good candidate to be analytically continued until $d=4$ as suggested by Weinberg. The natural candidate is instead the formulation based on the ${\rm Diff}$ symmetry group
with renormalization discussed in Eq.~\eqref{eq:martZanuEinstDiv}.

%%%%%%%%%%%%%%%%%%%%%%%%%%%%%%%%%%%%%%%%%%%%%%%%%%%%%%%%%%%%%
%%%%%%%%%%%%%%%%%%%%%%%%%%%%%%%%%%%%%%%%%%%%%%%%%%%%%%%%%%%%%
\section{Conclusions}
\label{sec:conclusions}
%%%%%%%%%%%%%%%%%%%%%%%%%%%%%%%%%%%%%%%%%%%%%%%%%%%%%%%%%%%%%
%%%%%%%%%%%%%%%%%%%%%%%%%%%%%%%%%%%%%%%%%%%%%%%%%%%%%%%%%%%%%

It is difficult to forecast the future of Weinberg's and Reuter's asymptotic safety program for quantum gravity. While we know very well that new computations --
increasingly technical and complex and, most likely, based on Wetterich's incarnation of the functional renormalization group method --
will probably further confirm the original conjecture by finding an appropriate ultraviolet fixed point, we must acknowledge that eventual scheme, gauge, or parametric dependences
of the results will not convince presently skeptical fellow scientists
to endorse the program.

In this chapter, we tried to give a path,
in part logical and in part historical,
that describes the onset of the original idea by Weinberg and its
developments. We covered mostly the early years, but also connected with recent research topics.
We purposedly skewed our narrative towards the perturbative results, which can explicitly be shown to be independent by all unwanted parameters, including the gauge ones, at least at the leading order in the expansion in $d=2+\varepsilon$
dimensions.

Our hope is that a renewed interest towards the perturbative perspective
in $d=2+\varepsilon$ will go hand-in-hand with the future nonperturbative developments in $d=4$. In fact, each approach has something to teach us and
both can learn from each other.
At the cost of sounding pretentious,
we could outline two important lessons coming from the perturbative approach
(among the many) that should be considered on the nonperturbative side.

The first and, to an extent, surprisingly trivial lesson
is that it is necessary to go on-shell
in order to show the gauge-independence of the final results,
which is a well-known fact of quantum field theory.
This is a step
that is, however, not as easy to take in the current applications of the functional renormalization group approach, considering also that few attempts have ever been made (see, for example, Ref.~\cite{Benedetti:2011ct, Baldazzi:2021orb, Baldazzi:2021ydj}). This step could have notable consequences when determining, for example, the dimensionality of the ultraviolet critical surface (i.e., the predictive power of the theory) in terms of the number of scaling operators that \emph{actually} exist on-shell and not only off-shell.

A second lesson comes from the fact that dimensional regularization requires that the dimensionality is continued below $d=2$ for Feynman diagrams to be finite,
but the results must be continued above $d=2$ in order to apply to the physical four-dimensional case.
This is strictly related to the problem of the (perturbative) instability of the conformal mode, and we recommend the general discussion of Ref.~\cite{Lauscher:2000ux} for some insights and implications.
It is known from simpler field theories (see, for example, Ref.~\cite{Giombi:2019upv}) that this type of continuation
can result in strong instanton contributions that cannot be neglected in the nonperturbative definition of the path-integral,
and has the potential side-effect of making the theory nonunitary.
In fact, unitarity itself is an often forgotten issue of the asymptotic safety proposal, although some work has been done in that direction \cite{Arici:2017whq,Nink:2015lmq}.

Important toy models that have the potential to teach us how asymptotically safe quantum gravity should behave above $d=2$ are $SU(N)$ Yang-Mills theories above $d=4$ (see an astute footnote of Ref.~\cite{Peskin:1980ay}). In fact, Yang-Mills gauge theories are asymptotically free in $d=4$
for some values of the parameters (like gravity in $d=2$),
so they are expected to be asymptotically safe above $d=4$ (like gravity is supposed to be above $d=2$). The case of $SU(2)$ Yang-Mills
in five dimensions has been studied in the past and recently,
both on the lattice \cite{Kawai:1992um,Florio:2021uoz,zippo},
where there is evidence only of a first-order phase transition
(which in layman terms means bad luck: no ultraviolet fixed point!)
and with the resummation of the perturbative series \cite{Morris:2004mg,DeCesare:2021pfb,zippo},
where errorbars caused by nonperturbative effects become quite considerable
in $d=5$. Conversely, the application of the functional renormalization group with the background field method undoubtedly suggests that there is a nontrivial ultraviolet completion \cite{Gies:2003ic}.
In short, regarding especially the change with the dimensionality of the nature of the UV behaviour,
the only evidence for local gauge theories that become asymptotically safe (above the dimensionality in which they are asymptotically free) is based on functional renormalization group methods.\footnote{%
This is not in contradiction with the existence of asymptotically safe gauge Yukawa theories in four dimensions.
}
This is certainly a point that deserves a careful consideration in the future.

As a final remark, we point out the importance of the definition of the path-integral. Several elements have to be considered when searching for the proper quantum theory of gravity, ultimately corresponding to a choice of functional measure in the partition function. Two elements that often reflect such (implicit) choice are the symmetry group and the parametrization of the fluctuations. In the case of quantum gravity and asymptotic safety, a fundamental question concerns the role of Weyl symmetry and, potentially, its anomaly. However, we realized how requiring the theory to be defined by an analytic continuation in the dimension $d$ constrains some of these aspects, mostly because of the different realizations of conformal (classical) theories in specific dimensions. Of course, a different approach as the functional renormalization group might restrict the analysis to a fixed dimensionality and give an entirely different point of view on the four-dimensional conformal anomaly.

\setcounter{section}{0}
\renewcommand{\thesection}{Appendix~\Alph{section}}
\renewcommand{\thesubsection}{\Alph{section}.\arabic{subsection}}

%%%%%%%%%%%%%%%%%%%%%%%%%%%%%%%%%%%%%%%%%%%%%%%%%%%%%%%%%%%%%
%%%%%%%%%%%%%%%%%%%%%%%%%%%%%%%%%%%%%%%%%%%%%%%%%%%%%%%%%%%%%
\section{Background field method}
\label{app:bkgFieldMeth}
%%%%%%%%%%%%%%%%%%%%%%%%%%%%%%%%%%%%%%%%%%%%%%%%%%%%%%%%%%%%%
%%%%%%%%%%%%%%%%%%%%%%%%%%%%%%%%%%%%%%%%%%%%%%%%%%%%%%%%%%%%%

We present a brief and informal outline of the background field method that summarizes the most important fetures of its application and the motivations behind it.
The standard path-integral of an (Euclidean) field theory is
based on the source-dependent functionals
\begin{eqnarray}\label{eq:pi-ZW}
 Z[J] &=& {\rm e}^{W[J]} = \int D\phi \,{\rm e}^{-S[\phi]+J\cdot \phi}\,,
\end{eqnarray}
where $J$ is the source for the field $\phi$ and the internal product consists in a sum over all internal indices and a covariant integration over spacetime.
The effective action is a functional of $\overline{\phi}=\langle\phi\rangle_J$, compactly defined as
\begin{eqnarray}\label{eq:pi-Gamma}
 {\rm e}^{-\Gamma[\overline{\phi}]} &=&  \int D\phi \,{\rm e}^{-S[\phi]
 +\frac{\delta\Gamma}{\delta\overline{\phi}}\cdot\left(\phi-\overline{\phi}\right)}\,,
 \qquad \overline{\phi}= \frac{\delta W}{\delta J}\,.
\end{eqnarray}
In the background approach one follows the strategy of decomposing the field $\phi$ into a background $\varphi$ and fluctuations $\chi$, with the aim of integrating over $\chi$.
The simplest choice is the linear split, $\phi = {\varphi} + \chi$,
though nonlinear implementations might be useful in some circumstance.
The background version of \eqref{eq:pi-ZW} is
\begin{eqnarray}\label{eq:background-ZW}
 Z_{\varphi}[J] &=& {\rm e}^{W_{\varphi}[J]} = \int D\chi \,{\rm e}^{-S[\varphi+\chi]+J\cdot \chi}\,,
\end{eqnarray}
where the dependence on the background $\varphi$ is kept parametrically. 
The background effective action is
\begin{eqnarray}\label{eq:background-Gamma}
 {\rm e}^{-\Gamma_{\varphi}[\overline{\chi}]} &=&  \int D\chi \,{\rm e}^{-S[\varphi+\chi]+\frac{\delta\Gamma_{\varphi}}{\delta\overline{\chi}}\cdot \left(\chi-\overline{\chi}\right)}\,,
 \qquad \overline{\chi} = \frac{\delta W_{\varphi}}{\delta J}\,.
\end{eqnarray}

There are two intimately connected reasons to apply the background field method.
On the one hand, we can consider the ``zero-point function'', i.e., $\overline{\chi}=0$,
and obtain a functional $\overline{\Gamma}[\varphi] \equiv \Gamma_{\varphi}[\overline{\chi}=0]$ that depends on a single field, which we argue briefly is (related to) the traditional effective action $\Gamma[\overline{\phi}]$.
On the other hand, the background field method allows us to preserve any symmetry, including nonlinear and gauge ones, throughout the computation in some form.

To argue the first point, consider the ``split'' symmetry induced by the arbitrary split $\phi = {\varphi} + \chi$:
\begin{eqnarray}\label{eq:split}
 \varphi \to \varphi - B\,, \qquad \qquad \chi \to \chi + B\,.
\end{eqnarray}
By construction $S[\varphi+\chi]$ is invariant under this transformation. Formally,
the original action satisfies the N\"other identity
$
 \frac{\delta S}{\delta\varphi} = \frac{\delta S}{\delta \chi}
$
implying that it is a function of the sum, $\phi = {\varphi} + \chi$.
Using \eqref{eq:background-Gamma}, one derives the split Ward identities
\begin{eqnarray}\label{eq:split-WI}
  \frac{\delta \Gamma_{\varphi}}{\delta \varphi}
  = \frac{\delta \Gamma_{\varphi}}{\delta \overline{\chi}}
  +{\cal A}\,,
\end{eqnarray}
where ${\cal A} \propto \langle \delta \log D\chi\rangle$ is some eventual anomaly induced by the noninvariance of $D\chi$ under \eqref{eq:split}
(further contributions are present for nonlinear splits generalizing \eqref{eq:split}).
Dimensional regularization is a procedure such that oftentimes ${\cal A}=0$,
so $\Gamma_{\varphi}[\overline{\chi}]$ must be a function of the sum, which can be denoted $\overline{\phi}=\varphi+\overline{\chi}$. As a consequence the (zero-point function of the) background effective action and the original effective action are the same
\begin{eqnarray}
 \overline{\Gamma}[\overline{\phi}] = \Gamma[\overline{\phi}]\,.
\end{eqnarray}

As for the second point concerning the preservation of symmetries, suppose that $\phi\to \phi'=F[\phi]$ is nonanomalous gauge and/or nonlinear symmetry of the action, $S[\phi] = S[\phi']$.
We deduce that it must be a symmetry of the effective action of \eqref{eq:pi-Gamma}, however, using standard Feynman diagrams and perturbation theory, it might not be obvious how to make it manifest at any stage of the computation.
In the background path-integral \eqref{eq:background-Gamma} the field is split, so there are two ``natural'' realizations of the symmetry.
The \emph{background symmetry}, in which the background changes as would the original field, while the fluctuation $\chi$ transforms linearly (in $\chi$ itself).
The \emph{full (quantum) symmetry}, in which the background remains invariant and the fluctuation takes over the full transformation originally belonging to $\phi$ (that is, generally, nonlinear in the fluctuation itself).

The background symmetry can be realized manifestly when covariantly expanding the action of any field (some attention must be given to nonlinear symmetries, but it is otherwise simple and straightforward).
The full quantum symmetry, on the other hand, is the physically relevant symmetry (e.g., the one that has to be gauge-fixed), because the background field is purely a computational device and should not transform.
However, the split Ward identities \eqref{eq:split-WI}, if satisfied and nonanomalous, allow to change the relative contributions of background and fluctuations. As a consequence, it is simple to deduce that, if the background \emph{and} split symmetries are satisfied, then the full quantum symmetry must also be satisfied!

%%%%%%%%%%%%%%%%%%%%%%%%%%%%%%%%%%%%%%%%%%%%%%%%%%%%%%%%%%%%%
%%%%%%%%%%%%%%%%%%%%%%%%%%%%%%%%%%%%%%%%%%%%%%%%%%%%%%%%%%%%%
\subsection{Background field: the case of metric gravity}
%%%%%%%%%%%%%%%%%%%%%%%%%%%%%%%%%%%%%%%%%%%%%%%%%%%%%%%%%%%%%
%%%%%%%%%%%%%%%%%%%%%%%%%%%%%%%%%%%%%%%%%%%%%%%%%%%%%%%%%%%%%

In theories with a dynamical metric $g_{\mu\nu}$ the relevant symmetry to consider is diffeomorphism invariance. Infinitesimally, a diffeomorphism is parametrized by a tangent vector field $\xi^\mu$ and the metric transforms as $\delta_\xi g_{\mu\nu} = 2\nabla_{(\mu} \xi_{\nu)}$, where the index of $\xi^\mu$ is lowered by the metric $g_{\mu\nu}$. If we perform a simple split over a background $\overline{g}_{\mu\nu}$, we have that 
$g_{\mu\nu}=\overline{g}_{\mu\nu}+h_{\mu\nu}$, where the fluctuations $h_{\mu\nu}$ are integrated over in the path-integral.
With this procedure, we have the identifications $\phi \to g_{\mu\nu}$ and $\chi \to h_{\mu\nu}$ of the previous section; the effective action is thus expected to be a functional of $\overline{\chi} \to \langle h_{\mu\nu} \rangle$ that depends parametrically on the background,
denoted $\Gamma_{\overline{g}}[\langle h\rangle]$.
Notice that $\Gamma_{\overline{g}}[\langle h\rangle]$ can be understood as a ``bimetric'' functional $\Gamma[\langle g\rangle,\overline{g}]
= \Gamma[\overline{g}+\langle h\rangle,\overline{g}]$, as was occasionally done in the asymptotic safety literature.

The \emph{background} transformations that are preserved through the quantization procedure are
\begin{equation}
\begin{split}
 \delta^b_\xi \overline{g}_{\mu\nu} = g_{\rho(\mu}\overline{\nabla}_{\nu)} \xi^\rho\,,
 \qquad
 \delta^b_\xi h_{\mu\nu} = \xi^\rho \overline{\nabla}_\rho h_{\mu\nu}
 + 2 h_{\rho(\mu}\overline{\nabla}_{\nu)} \xi^\rho\,,
\end{split}
\end{equation}
where we used the symmetric background connection $\overline{\nabla}_\rho\overline{g}_{\mu\nu}=0$.
The \emph{full quantum} transformation is instead
a complete diffeomorphism at fixed background,
$\delta_\xi \overline{g}_{\mu\nu} = 0$,
which is thus nonlinear in $h_{\mu\nu}$
\begin{equation}
\begin{split}
 \delta_\xi h_{\mu\nu} = g_{\rho(\mu}\nabla_{\nu)} \xi^\rho
 =
  \overline{g}_{\rho(\mu}\overline{\nabla}_{\nu)} \xi^\rho
  +\xi^\rho \overline{\nabla}_\rho h_{\mu\nu}
 + 2 h_{\rho(\mu}\overline{\nabla}_{\nu)} \xi^\rho
 +{\cal O}(h^2)
 \,,
\end{split}
\end{equation}
where ${\cal O}(h^2)$ hides computable, but more complicate, higher order contributions.
Nonlinear generalizations of the split do not affect the background transformation
of the background metric, but they do add further nonlinearities to the transformations of $h_{\mu\nu}$. These, however, can also be computed iteratively
without much difficulty.

The advantage of the background field method, assuming
that the split symmetry is not anomalous, is that computations of the effective action $\overline{\Gamma}[\overline{g}_{\mu\nu}]=\Gamma_{\overline{g}}[\langle h_{\mu\nu}\rangle=0]$
are performed in the limit $\langle h_{\mu\nu}\rangle=0$, so, at the leading order,
transformations of $h_{\mu\nu}$ can be largely ignored (although they do become important at next-to-leading order and, in general, whenever higher vertices with external $h_{\mu\nu}$ lines are needed). It is straightforward to see that
$\delta_\xi h_{\mu\nu}|_{h=0}= \delta^b_\xi \overline{g}_{\mu\nu}$, which is another way to see that the single-field effective $\overline{\Gamma}[\overline{g}_{\mu\nu}]$ is expected to be diffeomorphism invariant when split symmetry is not anomalous.

%%%%%%%%%%%%%%%%%%%%%%%%%%%%%%%%%%%%%%%%%%%%%%%%%%%%%%%%%%%%%
%%%%%%%%%%%%%%%%%%%%%%%%%%%%%%%%%%%%%%%%%%%%%%%%%%%%%%%%%%%%%
\section{Covariant computations and the heat kernel}
\label{app:covComputationsHK}
%%%%%%%%%%%%%%%%%%%%%%%%%%%%%%%%%%%%%%%%%%%%%%%%%%%%%%%%%%%%%
%%%%%%%%%%%%%%%%%%%%%%%%%%%%%%%%%%%%%%%%%%%%%%%%%%%%%%%%%%%%%

The heat kernel method allows to perform covariant computations of effective actions and Feynman-like diagrams in curved space. Here ``covariance'' refers
to both general covariance, as in General Relativity, and gauge covariance under eventual internal symmetry groups. The heat kernel method is a natural approach
to actual computations which are set up by the background field method discussed in the previous appendix.

Rather generally, the application of the background field method on an action $S[\phi]$ results in a Taylor-like expansion $S[\phi]=S[\varphi]+S_1[\varphi,\chi]+S_2[\varphi,\chi]+\cdots$. In many cases of interest, the quadratic part can be written
as $S_2[\varphi,\chi] =\frac{1}{2} \chi \cdot {\cal O} \cdot \chi$, where
we defined an operator of Laplace-type acting on the field's bundle
\begin{equation}\label{eq:operator-form}
\begin{split}
 {\cal O} = - g^{\mu\nu}\nabla_\mu\nabla_\nu + E\,,
\end{split}
\end{equation}
where $E=E(x)$ is a local endomorphism. In most other cases, kernels which differ from \eqref{eq:operator-form} can also be cast in the above form (for example, the square of a Dirac operator in curved space, $\slashed{\nabla}$, is $\slashed{\nabla}^2=\nabla^2 + \frac{R}{4}$). Everything depends on background fields from now on.

The Green function $G(x,x')$ of \eqref{eq:operator-form}
is formally defined as the inverse of ${\cal O}$
and is the basic building-block of covariant Feynman diagrams.
It solves the equation
\begin{equation}
\begin{split}
 {\cal O}_x G(x,x') = \delta^{(d)}(x,x')\,,
\end{split}
\end{equation}
where the $\delta$-function is a bi-scalar density such that $\int \!\!\sqrt{g}\,d^dx' \delta^{(d)}(x,x') f(x')=f(x)$. Covariant Feynman diagrams are products of Green functions and vertices acting as differential operators on them. Traditional Feynman diagrams can be seen as the momentum space representations of these products in flat space.

The heat kernel function of \eqref{eq:operator-form} is defined as the solution of the diffusion equation
\begin{equation}\label{eq:hk-def}
\begin{split}
%  &
 \partial_s {\cal G}(s;x,x') + {\cal O}_x {\cal G}(s;x,x') = 0\,,
%  \\&
\qquad
 {\cal G}(0;x,x') = \delta^{(d)}(x,x')\,.
\end{split}
\end{equation}
The formal solution is the exponential,
${\cal G}(s;x,x') = \langle x' | ~ {\rm e}^{-s {\cal O}} ~ | x\, \rangle$,
which can be easily related to the Green function by integrating over the diffusion ``time'' $s$
\begin{equation}\label{eq:green-sdw-relation}
\begin{split}
  G(x,x') = \int_0^{\infty} {\rm d}s \, {\cal G}(s;x,x') \,.
\end{split}
\end{equation}
Given that there is a practical way to compute ${\cal G}(s;x,x')$, given below, Eq.~\eqref{eq:green-sdw-relation} should be taken as the operative definition of $G(x,x')$.

The heat kernel has an asymptotic expansion for small values of the diffusion ``time'' $s$, known as the Seeley-de Witt expansion, which can be computed iteratively. To justify its form, consider first the solution of \eqref{eq:hk-def} in flat space and for $E=0$
\begin{equation}\label{eq:sdw-expansion-flat}
\begin{split}
 {\cal G}(s;x,x') &= \frac{1}{(4\pi s)^{d/2}}  ~ {\rm e}^{-\frac{\left|x-x'\right|^2}{4s}} \,.
\end{split}
\end{equation}
The flat space formula is generally ``covariantized'' as
\begin{equation}\label{eq:sdw-expansion}
\begin{split}
 {\cal G}(s;x,x') &= \frac{\Delta(x,x')^{1/2}}{(4\pi s)^{d/2}}  ~ {\rm e}^{-\frac{\sigma(x,x')}{2s}} \sum_{k\geq 0} a_k(x,x') ~ s^k\,.
\end{split}
\end{equation}
where the Synge-de Witt's world function, denoted $\sigma(x,x')$,
is half of the square of the geodesic distance between $x$ and $x'$,
and $\Delta(x,x')=(g(x) g(x'))^{-1/2} \det \left(-\partial_\mu\partial_{\nu'} \sigma \right)$ is the van Vleck determinant (it is a conventional normalization).
In flat space, $2\sigma(x,x')=\left|x-x'\right|^2$ and $\Delta=1$.
The bitensors $a_k(x,x')$ are the coefficients of the asymptotic expansion and contain the geometrical information of the operator ${\cal O}$ in terms of curvatures of the connection and interactions such as $E$.

Ultraviolet properties are local in meaningful and renormalizable quantum field theories. In terms of the Green function locality corresponds to $x\sim x'$,
formalized by the coincidence limit $x\to x'$.
Given a bitensor $B(x,x')$, the coincidence limit is denoted
$ [B] = \lim_{x'\to x} B(x,x') $ and regarded as a function of $x$.

The coincidence limits of (the derivatives of) the bitensors $\sigma(x,x')$, $\Delta(x,x')$ and $a_k(x,x')$ can be computed algorithmically
starting from few ``crucial'' equations. The first two are $\partial_\mu\sigma \partial^\mu \sigma = 2 \sigma$ and $\Delta^{1/2} \nabla^2 \sigma +2 \sigma^\mu \partial_\mu \Delta^{1/2} = d \Delta^{1/2}$, while the equations for $a_k$ can be derived by inserting \eqref{eq:sdw-expansion} in \eqref{eq:hk-def}
and expanding in powers of $s$. The boundary conditions are $[\sigma]=0$, $[\Delta^{1/2}]=1$, and $a_0(x,x')=1$ (the identity in the internal space). The coincidence limits relevant for \eqref{eq:operator-form} can be arranged by counting the number of covariant derivatives and curvatures, owing to the fact that $s$ is a dimensionful parameter.
They are known to a rather high order.

Using the heat kernel, it is often straightforward to compute the leading one-loop quantum corrections to the effective actions, which are expressed in terms of the trace-log formula
\begin{equation}\label{eq:trace-log}
\begin{split}
  \frac{1}{2}\Tr \log {\cal O} &= -\frac{1}{2}\int d^dx \sqrt{g} \int \frac{ds}{s} \, {\tr}\,  {\cal G}(s;x,x) \\
  &\overset{\rm IR\, reg}{\to} -\frac{1}{2}\sum_{k\geq 0}\frac{1}{(4\pi)^{d/2}}\int d^dx \sqrt{g} \int  \frac{ds}{s^{d/2-1+k}} \, e^{-sm^2} \tr [a_k]
  \,,
\end{split}
\end{equation}
which is true modulo field-independent constants. It is clear that \eqref{eq:trace-log} picks up only coincidence limits of the expansion of ${\cal G}(s;x,x')$,
that is, the coefficients $[a_k]$ upon using the Seeley-de Witt expansion.
Each integral over $s$ is generally doubly divergent for massless theories: UV divergences appear for $s\to 0$ and must be subtracted by renormalization, while IR divergences appear for $s\to \infty$ and can be regulated by adding a small mass as $e^{-sm^2}$ inside \eqref{eq:trace-log}. In dimensional regularization, the limit $m^2\to 0$ can be obtained by opportunely continuing $d$ in each term. In the simplest applications, minimal subtraction selects only one coincidence limit for a given ``critical'' $d$. For example, when expanding around the massless theories, $[a_1]$ is the coefficient of the dimensional pole in two-dimensional theories, while $[a_2]$ is the coefficient for four-dimensional ones (modulo normalizations).

For subleading computations, the ability to handle products of Green functions is paramount.
The expansion \eqref{eq:sdw-expansion} induces naturally an expansion of the Green function in \eqref{eq:green-sdw-relation}
\begin{equation}\label{eq:green-sdw-expansion}
\begin{split}
  G(x,x') = \sum_{k\geq 0} G_k(x,x') a_k(x,x')\,.
\end{split}
\end{equation}
The leading $G_0(x,x')$ and the subleading $G_k(x,x')$ for $k\geq 1$ are bilocal contributions to the Green function and
are determined by a simple integration over the heat kernel parameter $s$.
For example, assuming that $d$ is arbitrary (i.e., not even)
\begin{equation}\label{eq:Gk-definition}
\begin{split}
  G_k(x,x') = \frac{2^{d-2-2k}}{(4\pi)^{d/2}}
  \frac{\Delta^{1/2}}{(2\sigma)^{d/2-1-k}} \Gamma\left(\frac{d}{2}-1-k\right)
  \,.
\end{split}
\end{equation}
If $d$ is even one must account for logarithmic contributions of the form $\sigma^n\log \sigma$ modifying some order in the $k$ expansion above
(as seen in the Hadamardt representation of $G$).
When $d\sim 2$ there is an IR $\epsilon$-pole to subtract in the leading part
\begin{equation}\label{eq:Gk-definition-2d}
\begin{split}
  G_{0}(x,x')|_{d=2-\epsilon} =
  \frac{\Delta^{1/2}}{(4\pi)^{d/2}} \frac{\Gamma\left(d/2-1\right)}{(2\sigma)^{d/2-1}} +  \mu^{-\epsilon} ~ \frac{ \Delta^{1/2}}{2 \pi \epsilon}
  \,.
\end{split}
\end{equation}
If $d \sim 4$, the logarithm is in the subleading part
\begin{equation}\label{eq:G1-definition-4d}
\begin{split}
  G_{1}(x,x')|_{d=4-\epsilon} =
  \frac{\Delta^{1/2}}{(4\pi)^{d/2}} \frac{\Gamma\left(d/2-2\right)}{(2\sigma)^{d/2-2}} +  \mu^{-\epsilon} ~ \frac{ \Delta^{1/2}}{8 \pi^2 \epsilon} 
  \,.
\end{split}
\end{equation}
The scale $\mu$ is the reference scale of dimensional regularization.
The subtractions are the covariant analog of the infrared poles appearing in dimensionally regulated momentum-space Feynman diagrams.

%%%%%%%%%%%%%%%%%%%%%%%%%%%%%%%%%%%%%%%%%%%%%%%%%%%%%%%%%%%%%
%%%%%%%%%%%%%%%%%%%%%%%%%%%%%%%%%%%%%%%%%%%%%%%%%%%%%%%%%%%%%
\subsection{Covariant Feynman diagrams}
%%%%%%%%%%%%%%%%%%%%%%%%%%%%%%%%%%%%%%%%%%%%%%%%%%%%%%%%%%%%%
%%%%%%%%%%%%%%%%%%%%%%%%%%%%%%%%%%%%%%%%%%%%%%%%%%%%%%%%%%%%%

Leading quantum corrections at one-loop are generally covered by the simple use of \eqref{eq:trace-log} for as long as the kernel of the quadratic part of the bare action can be written in terms of a Laplace-type operator using the background method.
However, computations of subleading corrections (or corrections to composite operators), generally involve more complicate structures that can be diagrammatically summarized as covariant Feynman diagrams, which can be seen as the coordinate representation of the standard diagrams in momentum space \cite{Jack:1983sk}. In this appendix, we follow the presentation of Appendix~C of Ref.~\cite{Martini:2018ska} that summarizes some of the most important points of Ref.~\cite{Jack:1983sk}.

In practical applications, covariant Feynman diagrams are constructed as products of Green functions
and the vertices act as multilocal differential operators on them.
Using the case $d\sim 4$ as example, it is convenient to rewrite \eqref{eq:green-sdw-expansion} as
\begin{equation}\label{eq:green-sdw-expansion-rewritten}
\begin{split}
  G(x,x') = G_0(x,x') + G_1(x,x') a_1(x,x') + \overline{G}(x,x')\,,
\end{split}
\end{equation}
which isolates the first two contributions as potential sources of divergences in
the limit $x\sim x'$ from the rest, denoted $\overline{G}(x,x')$, which is regular in the limit. Beyond one-loop, diagrams will contain several structures:
local poles of the form $\frac{1}{\epsilon}$ that contribute to the renormalization group through beta functions, higher order poles up to the form $\frac{1}{\epsilon^L}$ for $L\geq 2$ at the $L$-th loop order,
and divergences multiplying the regular nonlocal part $\overline{G}(x,x')$,
besides finite nonlocal structures that depend on $\overline{G}(x,x')$.
In meaningful renormalizable field theories the local divergences must be subtracted, but only the leading $\frac{1}{\epsilon}$ one contributes to the
renormalization group running.\footnote{%
For example, two-loop divergences can only appear in the combination $\frac{\mu^{-2\epsilon}}{\epsilon^2}-2\frac{\mu^{-\epsilon}}{\epsilon^2}$, where the first terms is a genuine two-loop divergence, while the second comes from a one-loop divergence multiplying the one-loop counterterm. The logarithmic derivative with respect to $\mu$ is insensitive to this structure, in fact the divergences cancel in the limit $\epsilon\to 0$.
}
The divergences multiplying nonlocal structures of 
$\overline{G}(x,x')$ must cancel, or else the theory is nonlocal in the ultraviolet
(i.e., nonrenormalizable), which is often an important and nontrivial check
of the regularization process.

Using \eqref{eq:green-sdw-expansion-rewritten}, we can see that building blocks of the Feynman diagrams are structures of the form
$Q(x,x')\sigma(x,x')^{-b}$, where $Q(x,x')$ depends on the vertices of the theory
and $b$ on the number of propagators.
The covariant generalization of dimensional regularization starts with the basic relation
\begin{equation}\label{eq:basic-divergence}
\begin{split}
  \frac{1}{\sigma(x,x')^{\frac{d}{2}-c\epsilon}} \sim  \frac{(2\pi)^{\frac{d}{2}}}{c ~\epsilon ~ \Gamma(d/2)} ~ \mu^{-2c\epsilon} ~\delta^{(d)}(x,x')\,,
\end{split}
\end{equation}
which establishes equivalence for $x\sim x'$ and the reference scale $\mu$
to preserve the dimensionality (this relation can be proven using Riemann normal coordinates in curved space). In \eqref{eq:basic-divergence}, the dimension $d$ is the analytically continued dimension.

Higher inverse powers of the world function are also divergent.
They can be obtained using
\begin{equation}\label{eq:powers-of-sigma-relation}
\begin{split}
  & (\nabla^2-Y)\frac{\Delta^{1/2}}{\sigma^b} = b(2(b+1)-d) \frac{\Delta^{1/2}}{\sigma^{b+1}}\,,
  \quad
%   \\&
  Y(x,x')\equiv \Delta^{-1/2}\nabla^2\Delta^{1/2} \,,
\end{split}
\end{equation}
which can be proven easily applying the crucial relations.
For example
\begin{equation}\label{eq:basic-divergence+1}
\begin{split}
  \frac{\Delta^{1/2}}{\sigma(x,x')^{\frac{d}{2}+1-c\epsilon}} \sim  \frac{(2\pi)^{\frac{d}{2}}\mu^{-2c\epsilon}}{c ~\epsilon ~ d ~ \Gamma(d/2)} ~\left(\nabla^2 - \frac{R}{6}\right)\delta^{(d)}(x,x')\,,
\end{split}
\end{equation}
which is obtained inverting \eqref{eq:powers-of-sigma-relation} for $b=d/2$ and using the coincidence limits of the biscalars $[\Delta^{1/2}]=1$ and $[Y]=R/6$.

Whenever a bilocal tensor $Q(x,x')$ ``touches'' the Dirac delta on the right hand side of \eqref{eq:basic-divergence+1},
it can be replaced with its coincidence limit in the divergent part
$Q(x,x') \delta^{(d)}(x,x') = [Q]\delta^{(d)}(x,x')$.
If bilocal operators are separated
from the Dirac delta by covariant derivatives, it is necessary to integrate by parts.
For example,
we could manipulate as follows
\begin{equation}\label{eq:integration-by-parts}
\begin{split}
  &Q(x,x')\nabla_\mu \delta^{(d)}(x,x') = \nabla_\mu\left(Q(x,x') \delta^{(d)}(x,x')\right) - \nabla_\mu Q(x,x') \delta^{(d)}(x,x') \\
  & \sim \nabla_\mu\left([Q]\delta^{(d)}(x,x')\right) - [\nabla_\mu Q] \delta^{(d)}(x,x')\,.
\end{split}
\end{equation}
Similar manipulations can be performed for the cases with more derivatives.
Detailed applications of the heat kernel method at leading and nonleading orders
can be found in \cite{Jack:1983sk}.

\end{document}